\documentclass[a4paper]{article}
\pdfoutput=1
\usepackage{amsmath}
\usepackage{multicol}
\usepackage{color}
\usepackage{algorithmic}
\usepackage{graphicx}
\usepackage{amssymb}
\usepackage{algorithm}
\usepackage{arydshln}
\usepackage{upgreek}
\usepackage[left=2cm, right=2cm, top=2cm, bottom=2.5cm]{geometry}

\usepackage{xr}

\newcommand{\R}{\mathbb{R}}
\newcommand{\ha}{\hspace{-5pt}&\hspace{-5pt}}
\newcommand{\Nsamp}{N_{\mathrm{samp}}}
\renewcommand{\mp}{\hspace{-2pt}+\hspace{-2pt}}
\newcommand{\mm}{\hspace{-2pt}--\hspace{-2pt}}
\definecolor{blue}{RGB}{0,0,255}
\definecolor{red}{RGB}{255,0,0}

\begin{document}

\title{An algorithm for motif-based network design}


\author{Tuomo M\"aki-Marttunen$^{1,2}$\\
{$^1$Department of Signal Processing, Tampere University of Technology}\\
{$^2$Institute of Clinical Medicine, University of Oslo}\\
tuomo.maki-marttunen@tut.fi}

\maketitle

\noindent\textbf{Manuscript accepted for publication in IEEE/ACM Transactions on Computational Biology and Bioinformatics.}\\
\copyright \  2016 IEEE. Personal use of this material is permitted. Permission from IEEE must be obtained for all other uses, in any current or future media, including reprinting/republishing this material for advertising or promotional purposes, creating new collective works, for resale or redistribution to servers or lists, or reuse of any copyrighted component of this work in other works.

\vspace{20pt}

\begin{abstract}
A determinant property of the structure of a biological network is the distribution of its local connectivity patterns, i.e., \textit{network motifs}.
In this work, a method for creating directed, unweighted networks while promoting a certain combination of motifs is presented.
This motif-based network algorithm starts with an empty graph and randomly connects the nodes by advancing or discouraging the formation of chosen motifs.
The in- or out-degree distribution of the generated networks can be explicitly chosen.
The algorithm is shown to perform well in producing networks with high occurrences of the targeted motifs, both ones consisting of 3 nodes as well as
ones consisting of 4 nodes.
Moreover, the algorithm can also be tuned to bring about global network characteristics found in many natural networks, such as small-worldness and modularity.
\end{abstract}

\section{Introduction}

A prerequisite for understanding the function of a biological network is thorough knowledge on its structure and dynamics as well as an insight on their relationship \cite{sporns2011human,newman2003structure}.
Revealing structure-function relationships is an utterly challenging task due to the multitude of structural and dynamical measures involved \cite{strogatz2001exploring,rubinov2010complex}
and the correlations therein \cite{arenas2008synchronization,pernice2011how}.
This raises a need for considering a large diversity of network structures \cite{palla2010multifractal}, which can be computationally generated and used to test hypotheses and benchmark performance \cite{bois2015probabilistic}.
To date, a great number of network generation algorithms exist, but many of them are specific to the underlying field of study \cite{kitano1990designing,eubank2004modelling,robins2007recent,netmorph,acimovic2015effects,lu2009next}.

One of the most widely used algorithm is that of Watts and Strogatz (WS) \cite{watts1998collective}, which allows the generation of networks ranging from locally connected to random (in the Erd\H{o}s-R\'enyi
\cite{erdos1960evolution} sense), and the \textit{small-world networks} that lie in the middle of this range. Another widely used algorithm is the Barab\'asi-Albert network generation algorithm
\cite{barabasi1999emergence}, which allows one to create networks with power-law distributed degree using a preferential attachment rule. Both of these algorithms were designed for undirected
networks, but they can be extended to generate directed networks as well \cite{prettejohn2011methods}. Nevertheless, the variety of networks they produce is rather limited.

Many biological systems such as \textit{C. Elegans} connectome and gene regulatory networks possess distributions of local connectivity patterns, i.e. \textit{motifs}, that differ from those predicted by
  simple network models \cite{milo2002network}.
Thus, improved algorithms should generate alternative network models that take into account the distribution of motifs within the network.
In this work, an algorithm for randomly creating directed, unweighted networks with high occurrences of certain local connectivity patterns is presented and analyzed.
The strength of the motif-based analysis is that the network generation is guided by the distribution of the local connectivity patterns and the emergent
  global characteristics can hence be viewed as a combination of these basic building blocks of network structure.
This \textit{motif-based network} (MBN) design algorithm is effective in comparison to existing methods \cite{bois2015probabilistic} in producing networks with any of the local connectivity
  patterns overrepresented, but it is also capable of bringing up networks with certain global features found in many real-world networks, such as small-worldness \cite{watts1998collective}
  or division into communities \cite{newman2006modularity}.


\section{Methods}
The MBN algorithm starts with $N$ nodes and no connections between them. Let us denote the network connectivity matrix as $M\in\{0,1\}^{N\times N}$, where $M_{ij}$ stands for the
connection \textit{from} node $i$ \textit{to} node $j$. Self-connections are prohibited in this study. 
Each node is drawn a target number of inputs from the in-degree distribution, $n_i\sim p_{\mathrm{in}}$,
and the algorithm is then iterated until each node has been set the chosen number of inputs. At each iteration, the node that will be given an input is chosen by random.
Each possible input node --- that is, a node that does not yet project to the considered node --- is given points basing on how many motifs of
each kind would be formed and how many broken if the considered node was chosen as an input, and the node with highest score is picked.
A pseudo-code for the algorithm is given below, and a MATLAB implementation of the algorithm is publicly available at http://github.com/tuomomm/MBN/.
\begin{algorithmic}
\FOR {node index $i\in\{1,\ldots,N\}$}
 \STATE Draw number of inputs $n_i\sim p_{\mathrm{in}}$.
\ENDFOR
\WHILE {not all edges set}
 \STATE Count the number of unassigned inputs of each node: \\ \hspace{10pt}$\forall i$: $u_i\leftarrow n_i - \sum_{j=1}^NM_{ji}$.
 \STATE Randomly pick node $k$ from a weighted distribution $P(k)=u_k/\sum_{i=1}^Nu_i$.
 \STATE Count the points for possible inputs of node $k$ as \\ \hspace{10pt}$\lambda = \mathrm{Calculatepoints}(k,M,w)$
 \STATE Pick the highest scoring node $i = \mathrm{argmax}_{j\ \mathrm{s.t.}\ M_{jk}=0}\lambda_j$ (if multiple nodes with the highest score, pick one of them by random).
 \STATE Make a connection from $i$ to $k$, i.e., set $M_{ik}=1$.
\ENDWHILE
\end{algorithmic}

\begin{figure}[hb!]
\centering
\includegraphics[clip,trim=64pt 7pt 156pt 598pt, width=0.75\textwidth]{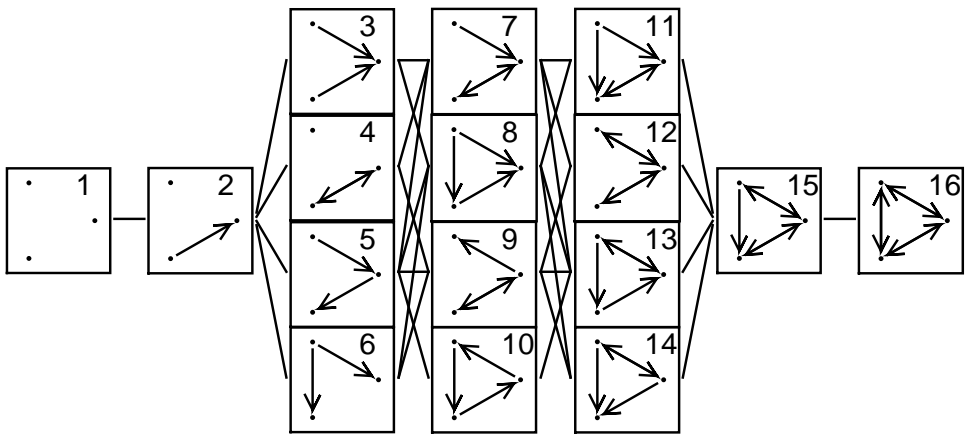}
\caption{The three-node motifs and their relations. A line between the motifs indicates that the left-hand motif can be derived from the right-hand motif by removing one edge.}
\label{fig1}
\end{figure}

The back-bone of the algorithm is the function Calculatepoints, which is given as arguments the target node $k$, the existing connectivity matrix $M$,
and the weight vector $w\in\R^{N_{\mathrm{mot}}}$ indicating the preference of different motifs.
$N_{\mathrm{mot}}$ is the number of possible motifs of the considered size: it is 16 in three-node motifs (see Figure \ref{fig1}) and 218 for four-node motifs.
The function goes through the existing connectivity patterns formed by target node $k$, each possible input node $i$, and other auxiliary nodes $j$ (in three-node motifs, one auxiliary node $j$ is
involved, while in four-node motifs two auxiliary nodes $j_1$ and $j_2$ are needed, etc.).
Let us call the motif formed by the nodes $i$, $j$ and $k$ at the time of picking an input a \textit{pre-motif}, with a distinction to the term \textit{motif} used elsewhere in this work
in that it is index-specific and used only for labeling the connectivity patterns prior to picking an input.
The function calculates the score $\lambda_i$ for each node $i\neq k$ that is not yet an input for node $k$ by calculating the numbers of pre-motifs and determining how many motifs of
each kind would be created and how many destroyed if an edge was drawn from $i$ to $k$.
For three-node motifs, the pseudo-code for the function is as follows:
\begin{algorithmic}
  \STATE $\lambda=$ Calculatepoints($k$,$M$,$w$):
  \FOR {node index $i\neq k$ s.t. $M_{ik}=0$}
   \FOR {$r\in\{1,\ldots,{N_{\mathrm{premot}}}\}$}
    \STATE Set the number of existing pre-motifs to zero:\\ \hspace{10pt}$Q_{ir}\leftarrow 0$
    \FOR {$j\in\{1,\ldots,N\}\backslash\{i,k\}$}
     \IF {the connections between nodes $i$, $j$, $k$ form the pre-motif $r$}
      \STATE Set $Q_{ir}\leftarrow Q_{ir}+1$
     \ENDIF
    \ENDFOR
   \ENDFOR
   \STATE Count the points for node $i$ as \\ \hspace{10pt}$\lambda_i \leftarrow\sum_{r=1}^{N_{\mathrm{premot}}}\sum_{m=1}^{N_{\mathrm{mot}}}Q_{ir}G_{rm}w_m$
  \ENDFOR
  \STATE Return vector $\lambda$
\end{algorithmic}
The matrix $G\in\R^{N_{\mathrm{premot}}\times N_{\mathrm{mot}}}$ is fixed such that the entry $G_{rl}$ indicates whether a motif $l$ is formed (1),
abolished (-1), or neither of the two (0) when an edge is added to a pre-motif of type $r$. See supplementary material Section \ref{sec_matrix_G} for the full
matrix $G$ and a visual presentation of the same data (Table \ref{AB3}).

The iterative nature of the MBN algorithm places a challenge for the formation of the highly connected motifs. As an example, 
in the case of promoting a fully connected motif, the algorithm only gives non-zero points if a next to fully connected motif (motif 15) exists in the network. 
Until such a motif is formed, the connections are drawn by random. 
To overcome this problem, the weights can be adapted such that while the motif $l$ is given full points ($\tilde w_l$),
all motifs that become motif $l$ by the addition of an edge are given an additional $\frac 1N\tilde w_l$ points, and all motifs that become motif $l$ by adding two edges are given
an additional $\beta\frac 1{N^2}\tilde w_l$ points, $\beta$ denoting the number of ways in which this can be attained, and so forth.
This way, the formation of the highly connected motifs is facilitated by encouraging the formation of the intermediate motifs.
Conversely, if a negative weight is given for some motifs, the formation of intermediate motifs is hindered as well.
The effective weights that are conveyed to the algorithm can be calculated as
\begin{equation}w = (I+\frac FN+\frac{F^2}{N^2}+...)\tilde w,\label{eq_adapt}\end{equation}
where $\tilde w\in\R^{N_{\mathrm{mot}}}$ are the weights
of the preferred motifs, $I$ is the identity matrix, and $F\in\R^{N_{\mathrm{mot}}\times N_{\mathrm{mot}}}$ is an adaptation matrix determined by the relations between the motifs
(see Supplementary material Section \ref{sec_matrix_F} for the adaptation matrix for three-node motifs).

\begin{figure}[hb!]
\centering
\includegraphics[clip,trim=82pt 154pt 90pt 470pt, width=0.99\textwidth]{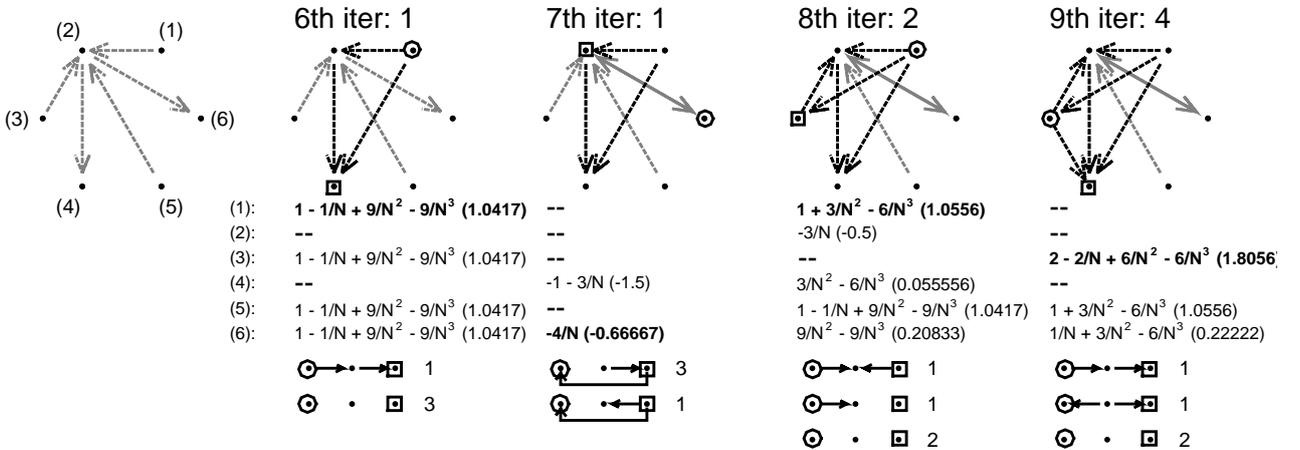}
\caption{Four iterations of the MBN algorithm pronouncing motif 8 (FF motif), network size $N=6$. The weights are chosen as $\tilde{w}_i=\delta_8(i)$,
and they are adapted using Eq. (\ref{eq_adapt}). The number of motifs of the specific type is announced beside the
iteration number. The bidirected edges are highlighted with solid lines, while
unidirected edges are dashed. Grey edges represent connections that are not part of an FF motif, while black edges contribute to at least one FF motif.
The square around a node marks the node $k$ for which an input was chosen at the considered iteration, and the circle denotes the node $i$ that was chosen as an input.
The in-degree distribution was binomial with $p=0.33$. Below each graph, the score of each node is given, a missing number indicating that the node is already an input for the considered node,
and below the scores, the pre-motifs that contributed to the points of the highest scoring node and their numbers are displayed. As an example, in the eighth iteration, one FF motif is created by the 
addition of the new edge, one intermediate motif of two edges (divergent motif, \#6) is created and one abolished (convergent motif, \#3), two intermediate motifs of one edge (motif \#2)
are created and one abolished, and two empty motifs (motif \#1) are abolished. Given the adaptation of the weights, this gives the score $1+\frac0N+\frac3{N^2}-\frac6{N^3}$.
}
\label{fig2}
\end{figure}

\subsection{Illustration of the MBN algorithm functioning in promoting three-node motifs}

Let us consider the promotion of the three-node feed-forward (FF) motif, which is motif 8 in Figure \ref{fig1}.
The motif can be promoted in the MBN framework using a preferred motif vector $\tilde w = \delta_8$, which by Eq. \ref{eq_adapt} gives an absolute weight vector
$w = \delta_8 + \frac 1N(\delta_3+\delta_5+\delta_6) + \frac 3{N^2}\delta_2 + \frac 3{N^3}\delta_1$.
Figure \ref{fig2} shows four iterations (sixth to ninth) of the MBN algorithm using this weight vector in a network of size $N=6$,
the points given for each node, and the present pre-motifs formed by target node (square) and the node chosen as input (circle) at each iteration.
At the sixth iteration, all possible input nodes have the exact same score (one FF motif formed and several intermediate motifs formed and destroyed),
and hence one of them (node 1) is chosen by random.
At the seventh iteration, both possible input nodes have a negative score, and hence the one with smaller absolute value --- the one that breaks
intermediate motifs but not the preferred motif --- is chosen.
At the eighth and ninth iteration, input nodes are chosen such that one and two preferred motifs are formed, respectively, and several intermediate
motifs formed and broken.

\section{Results}

\subsection{MBN algorithm succeeds in promoting a single motif}
\begin{figure}[hb!]
\centering
\includegraphics[clip,trim=10pt 125pt 250pt 368pt, width=0.7\textwidth]{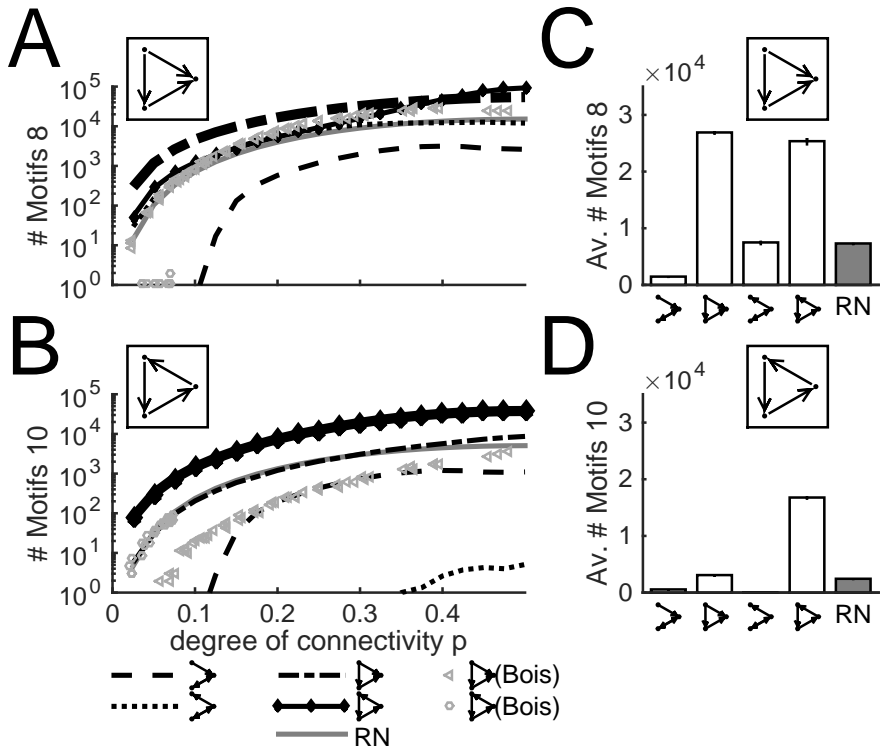}
\caption{\small The networks produced with the MBN algorithm pronouncing FF (motif 8) or FB motifs (motif 10) show a significant
increase in the numbers of the corresponding motif compared to other networks. 
\textbf{A, B}: The curves show the numbers of FF (A) or FB (B) motifs as a function of connection probability
$p$. Network size is $N=100$, and binomial in-degree distribution is used. The different curves show the numbers of the considered motif when different weights are used:
$\tilde w=\delta_7$,$\delta_8$,$\delta_9$,$\delta_{10}$, or $0\in\R^{16}$ (random network, denoted by RN). 
In both panels, the curve of the MBN promoting the considered motif is thickened. The values of the curves represent sample averages of $\Nsamp=200$ realizations.
\textbf{C, D}: The numbers of the three-node-three-edge motifs shown in (A) and (B), averaged (integrated and multiplied by two) over connection
probabilities $0<p\leq 0.5$. The tick on the bar shows the standard deviation of these average values, invisible ticks indicating very small deviations.
Comparisons are made among MBNs pronouncing another motifs with the same number of edges, and with random networks. The medians of the motif count distributions
are always significantly higher in those MBNs that pronounce the corresponding motif than in any other shown network (U-test, $p=0.05$).
}
\label{fig3}
\end{figure}

The performance of the algorithm in promoting three-node feed-forward and feed-back (FB) motifs (motifs 8 and 10 in Figure \ref{fig1}) in networks of size $N$=100 is
  illustrated in Figure \ref{fig3}. 
The preferred motifs are delta-weighted as $\tilde w_i = \delta_l(i)$, $l=7,8,9$ or $10$ i.e., in each network generation the formation of only one of the three-edge motifs is pronounced. 
The numbers of emerging FF and FB motifs are compared to the corresponding motif counts in random networks
  and networks produced by the MBN algorithm using a different weight vector.
The MBN focusing on the preferred motif repeatedly produces the greatest number of the preferred motif, except in the case of FF motif count in the dense networks,
  where the MBN promoting the FB motif outperforms the MBN promoting the FF motif.
This kind of interdependence in the motif counts is not unexpected, see Figure \ref{figS0} for an illustration on why this happens when promoting the FB motif.
Figure \ref{fig3} also shows that the algorithm performs well in comparison to the promotion of these motifs using the method of \cite{bois2015probabilistic}.
For this, the desired proportions of FB vs FF loops (parameters $u$ and $v$ in \cite{bois2015probabilistic}) were set 1:10000 (for promoting FF motifs) or 10000:1 (for promoting FB).
Nevertheless, in case of FB motif, this comparison may be unfair, as the method of \cite{bois2015probabilistic} considers cumulative motif occurrence: adding an edge to motif 10
  does not decrease the number of FB motifs in their framework. 
This is reflected on zero values of motif 10 counts in the experiment of Figure \ref{fig3}.
However, better performance could be expected by fine-tuning parameters $u$ and $v$.
The performance and computational load of the method of \cite{bois2015probabilistic} also depends on the number of iterations used.
The authors of \cite{bois2015probabilistic} suggest a use of up to 2 billion iterations for networks of size $N=423$ --- in Figure \ref{fig3}, 80 million iterations were used as
  larger numbers of iterations did not significantly improve the results.
A generation of network of size $N=100$ took 25.5$\pm$0.7 min using the algorithm of \cite{bois2015probabilistic}, while the MBN generation of the same size took
0.56$\pm$0.06 s ($p=0.025$) to 11.6$\pm$0.2 s ($p=0.5$).
The authors of \cite{bois2015probabilistic} stated a much smaller CPU time (only 2 minutes) for generating a network of size $N=423$ using optimized code and dual core computer.
The generation of an MBN of this size using single-core computers took 2.8$\pm$0.3 min ($p=0.025$) to 56.7$\pm$7.6 min ($p=0.5$).

In a similar fashion as in Figure \ref{fig3}, Figure \ref{figS1} shows the collection of data for all three-node motifs and confirms superiority of the MBN scheme in
  promoting a single network motif in comparison to random networks and other MBN instances, and Figure \ref{figS2} reproduces this in large ($N$=1000) networks.
Figure \ref{figS3} compares the results of Figure \ref{figS1} against those produced by MBN algorithm without the weight adaptation,
  and shows that the adaptation improves the promotion of the highly connected motifs.
The algorithm performance can also be validated against theoretical connectivity strategies that aim to maximize numbers different motifs.
In a general case this is extremely difficult, but for empty motifs such strategies can easily be defined, see Supplementary material Section \ref{sec_theoretical}.
Figure \ref{figSempty} shows that the MBN algorithm promoting empty three-node motifs performs on a level between two sub-optimal theoretical strategies.

To confirm that the method works when generalized to more complex local connectivity patterns as well,
  the performance of MBN algorithm in the case of four-node motif promotion is tested next.
Figure \ref{figS4} shows that the extension of MBN algorithm to four-node motifs performs well in a task similar to that of Figure \ref{fig3} and \ref{figS1}.
The body of the algorithm remains the same in the extension to motifs of four nodes, but the calculation of the points becomes more tedious (see the function Calculatepoints for four-node motifs
in Supplementary material Section \ref{sec_extension_to_four}).
As the maximal computational cost of the three-node motif algorithm is in the order of $O(EN^2)$ logical operations ($E$ denoting
the total number of edges), the cost of the four-node motif algorithm is in the order of $O(EN^3)$ operations.

\subsection{MBN algorithm allows promotion of global structural properties: Generation of small-world networks and networks with community structures}

\begin{figure}[hb!]
\includegraphics[clip,trim=0pt 0pt 30pt 350pt, width=0.99\textwidth]{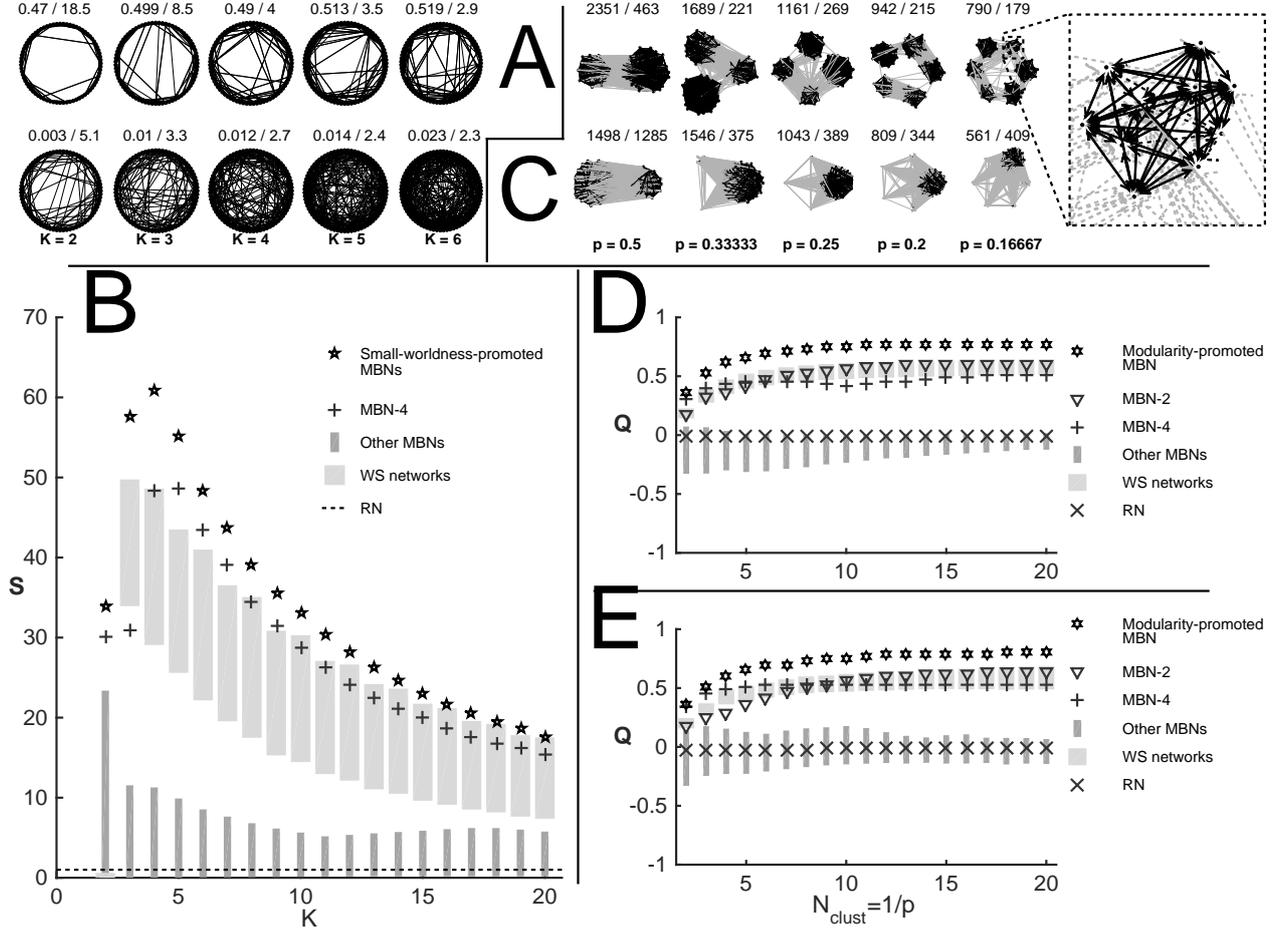}
\caption{\small MBN algorithm parameters can be tuned to promote global network features. \textbf{A}: Illustration on MBNs generated with weights $\tilde w=(-1.351,0,0,1.407,0,0,0,0,0,0,0,1.755,0,0,0,0.567)$
(upper panels), and the corresponding random networks (lower panels).
Network size is $N=75$, and the in-degree is distributed as $p_{\mathrm{in}}(d)=\delta_K(d)$, where the number of inputs ranges from $K=2$ (left) to $K=6$ (right).
The networks are illustrated using a scheme that iteratively permutes positions of the nodes to find the minimum summed distance between the linked nodes.
The clustering coefficient and the harmonic mean of the path lengths are printed for each network (CC / PL): The MBNs show greater clustering and longer path lengths than the random networks.
\textbf{B}: The small-worldness index of small-worldness-promoted MBNs (weights chosen as in (A)), delta-weighted MBNs, and WS networks as a function of number of inputs $K$.
The MBN with $\tilde w=\delta_4$ is plotted alone for its distinctive behaviour, while the range of average values of the other delta-weighted MBNs is shown with the shaded bars.
The dimly shaded bars show the range of small-worldness indices of WS networks with rewiring probability $q\in [0.0001,0.2]$ (the maximum small-worldness is attained on this range for all $K>2$).
The small-worldness-promoted MBNs gain large small-worldness indices compared to those of random networks and other MBNs, and even larger than those of the WS networks.
Network size is $N=200$, average taken over 200 networks.
\textbf{C}: Illustration on modularity-promoted MBNs, generated with weights 
$\tilde w=(1.852,1.3,0,0.838,0,0,0,0,0.084,0,0,-2.111,0,0,0,0.1317)$ (upper panels), and the corresponding
random networks (lower panels).
The networks shown are densely connected, $p$ ranging in inverted integers from $1/2$ (left, hierarchical clustering to two groups shown) to $1/6$  (right, hierarchical clustering to six groups shown).
The numbers of intra-cluster and inter-cluster edges (intra / inter) are shown for each network.
The clustering of modularity-promoted MBNs to the $1/p$ clusters shows both less deviation in cluster sizes and better ratio of intra vs. inter-cluster edges than the clustering of the corresponding random networks.
\textbf{D}: Modularities calculated for densely connected modularity-promoted MBNs (weights chosen as in (C)), delta-weighted MBNs, and WS networks with $q\in [0.0001,0.2]$ as a function
of number of clusters $N_{\mathrm{clust}}=1/p$.
The modularity-promoted MBNs show the highest values of modularity, followed by MBN-2, MBN-4, and WS networks in varying order.
Network size is $N=200$, sample size 200.
\textbf{E}: The result of (D) is reproduced with an alternative clustering algorithm \cite{zhou2005learning}.
}
\label{fig4}
\end{figure}

Contrary to Watts-Strogatz (WS) algorithm \cite{watts1998collective} and many follow-up network algorithms that first set spatial positions for the nodes and then assign the edges according to this topology,
the MBN algorithm focuses on the local connectivity patterns only, and the global structure emerges as a byproduct. In many cases, the global structure remains unidentifiable in the way that it
cannot be characterized as any easily recognizable topology. However, the weights can be chosen such that the generated networks possess certain global characteristics. Figure
\ref{fig4} shows how small-world networks and networks with community structure can be generated by choosing the MBN algorithm parameters carefully. For this ``careful'' choice, in this work,
a genetic algorithm is used to find optimal or sub-optimal weights.
During each evaluation of the optimized function, 20 networks of smaller size are generated for each desired in-degree distribution.
The function output is chosen as the average small-worldness index or the average modularity of the formed networks, and this function is maximized over possible values of weight $\tilde w$
as described below.

For the small-world networks, the optimization is carried out by maximizing \textbf{small-worldness} index, which is calculated as \cite{humphries2008network,prettejohn2011methods}
\begin{equation}S(M) = \frac{C(M)/C_{\mathrm{rand}}}{L(M)/L_{\mathrm{rand}}},\label{eq_sw}\end{equation}
where $C(M)$ is the clustering coefficient of the graph $M$ and $C_{\mathrm{rand}}$ is the mean clustering coefficient of a random graph with in-degree distribution equal
to that used for the generation of graph $M$. In a similar manner, $L(M)$ is the harmonic mean of lengths of all shortest paths in graph $M$ and $L_{\mathrm{rand}}$ is the corresponding value in a random graph.
The clustering coefficient of the graph $M$ is calculated as the average over local clustering coefficients, where the local clustering coefficient
of a node is defined as the number of connected triangles in its neighbourhood \cite{watts1998collective}.
For promoting the community structure, the measure of \textbf{modularity} is maximized. The modularity of a graph describes to what extent the connectivities within the clusters are denser than in a
graph where the edges are drawn at random \cite{newman2006modularity}. It is calculated as \cite{arenas2007size}
\begin{equation}Q(M) = \frac 1E\sum_{\mathrm{cluster}\ {\cal I}}\sum_{i,j\in {\cal I}}(M_{ij} - \frac{m_i^{\mathrm{out}}m_j^{\mathrm{in}}}E),
\label{modularity}\end{equation}
where $m_i^{\mathrm{out}}$ and $m_i^{\mathrm{in}}$ are the numbers of outputs and inputs, respectively, of node $i$.
In general, the modularity should be maximized over all possible partitionings, but this is an NP-hard problem \cite{arenas2007size}.
In this work, a widely accepted partitioning scheme, the hierarchical clustering \cite{girvan2002community}, is applied, using the Hamming distance in the output and input node patterns as the determining
factor for joining two clusters. For details on calculating the clustering coefficient and modularity, and on the parameter optimization,
see Supplemental material Sections \ref{sec_clustering_coefficient}--\ref{sec_parameter_optimization}.

Figure \ref{fig4}A illustrates the structure of the small-worldness-promoted MBNs, while Figure \ref{fig4}C illustrates the structure of the modularity-promoted MBNs.
Global structural features of these networks, namely, the tendency of forming many local connections and few long-range connections (A) or the tendency to form modules of
high intra-cluster connectivity (C), can be identified by choosing a proper placement of the nodes. Such features are not present in random networks with the corresponding in-degree distribution.
Figure \ref{fig4}B shows the small-worldness indices of the small-worldness-promoted MBNs, and their relation to the corresponding statistics in directed WS networks and other MBNs.
The small-worldness-promoted MBNs show constantly high small-worldness indices, comparable and even higher than those of directed WS networks, which are generated by first assigning each node
an input from its $K$ nearest neighbours in the ring and then randomly rewiring the source nodes of the edges with a rewiring probability $q$. While the small-worldness index of
a random network is 1 by definition, the delta-weighted MBNs ($\tilde w=\delta_i$, $i=1,\ldots,16$) possess small-worldness indices both below and above 1.

The values of modularity are shown for different densely connected networks in Figure \ref{fig4}D. The in-degree of these networks is binomially distributed with connection probability depending
on the targeted number of clusters as $p = 1/N_{\mathrm{clust}}$.
The modularity-promoted MBNs show prominent community structure: Their modularity exceeds the value 0.7 for $N_{\mathrm{clust}}\geq 10$, while the theoretical
maximum for the modularity, corresponding to a perfect community structure, is $1-\frac 1{N_{\mathrm{clust}}}$.
The directed WS networks, while exhibiting a strong neighbourhood
  structure, lack a division to strongly connected communities and hence show constantly lower values of modularity than the modularity-promoted MBNs.
The average modularity of random networks is approximately zero.
Qualitatively similar results are obtained by using an alternative partition method \cite{zhou2005learning} in the calculation of modularity, which is shown in Figure \ref{fig4}E.
Moreover, the results remain similar if a slightly simpler formula
of modularity, where it is approximated that $\frac{m_i^{\mathrm{out}}m_j^{\mathrm{in}}}E\approx p$ (cf. \cite{reichardt2004detecting}), is applied, as shown in Figure \ref{figS5}.

Figure \ref{fig4} shows that in addition to the small-worldness-promoted MBNs, MBNs with weight vector $\tilde w=\delta_4$ express high small-wordness indices. 
Similarly, MBNs with weight vector $\tilde w=\delta_2$ or $\tilde w=\delta_4$ are highly modular, although not as modular as modularity-promoted MBNs.
This interplay between motifs 2 and 4 and modularity index as well as that between motifs 4 and small-worldness is analyzed in Figure \ref{figS6} by considering a continuum of networks
  from single motif-promoting MBNs to the optimized MBNs. 
Although all networks show higher modularities (Figure \ref{figS6}A) or higher small-worldness indices (Figure \ref{figS6}B) than random networks,
  there is a small range of weight vectors localized around the optimized weights that excel in these measures across many connection probabilities.

\section{Conclusions and discussion}
The algorithm for motif-based network (MBN) generation was presented and analyzed. The algorithm performs well in generation of network structures
with high occurrences of chosen three- or four-node motifs. Moreover, it was shown that networks with higher-order structural properties, namely
small-worldness and modularity, are successfully generated using the MBN algorithm.

The strength of the MBN algorithm is that it is built upon controlling the occurrences of the basic building blocks of network connectivity, the motifs \cite{milo2002network}, in a way that also allows the emergence
of certain large-scale connectivity structures. This is an outstanding property with respect to many other network generation algorithms where the emerging connectivity patterns are dependent on the
initially determined locations of the nodes \cite{watts1998collective,netmorph,acimovic2015effects,itzhack2010random}. An important contribution was made by \cite{palla2010multifractal},
where an algorithm is presented for generating
a wide range of different networks using only two parameters. Nevertheless, the interpretation of these two parameters is extremely difficult in the terms of the resulting graph and its properties. The same can
be said of a recent algorithm \cite{lennartsson2012specnet} which, like the algorithm of \cite{palla2010multifractal}, first generates a probability measure for creating each of the links $i\rightarrow j$ basing
on few parameters, and then samples the whole network.
While the MBN algorithm does provide a meaningful interpretation of its model parameters, it has certain restrictions as well, one being its large computational cost,
  and another being the requirement of determining the network in-degree in advance.
The degree distribution is, however, a key property in all kinds of networks \cite{newman2003structure}, and is therefore often predetermined in network generation applications. 
Similar approach was chosen in \cite{serrano2005tuning}, where, in a undirected network framework, first a degree distribution is chosen, and the algorithm then optimizes the generated
network to have a chosen clustering coefficient.
Earlier work on generating networks with a chosen motif preference has been carried out in \cite{nykampgenerating} and \cite{bois2015probabilistic}, but using different premises.
In \cite{nykampgenerating}, the focus was set on motifs consisting of a chain of nodes and the possibility of the chain connecting into a loop was not considered, and
in \cite{bois2015probabilistic}, the motif occurrence was considered cumulative and the networks considered were probabilistic.
The MBN algorithm performance was compared against the latter and was found efficient both in terms of computation time and resulting numbers of motifs (see Figure \ref{fig3}A).

In the MBN algorithm, the choice of whether in- or out-degree is predetermined can be worked around by possible transposing of the final connectivity graph --- notice that the motifs for which the weights
are given have to be transposed as well. The restriction of network degree as well as the network size could be avoided by generating the network by iteratively adding nodes in a similar way as in
Barab\'asi-Albert \cite{barabasi1999emergence} networks (or their extension for directed networks \cite{bollobas2003directed}), but mimicking the input selection rules of MBN algorithm. However, this requires careful study of different attachment schemes and is therefore left for future work.

The work at hand concentrates on three-node and four-node motifs of uniform
networks, but the framework of the algorithm allows motif-based design of non-uniform
networks as well. 
This is particularly important in models of neuronal networks, where the nodes (neurons) are typically labelled either excitatory or inhibitory. 
The motifs in the MBN framework can be redefined to account for this at the expense of the computational load of the algorithm: 
In the case of three-node motifs, one would have to define the weights of 104 different motifs instead of the 16 motifs that exist in uniform networks.
This is due to the fact that each of the 16 motifs
can have either 4, 6, or 8 different instances when two node types are applied, depending on the level of symmetry of the motif. However, the adaptation matrix $F$ can be calculated in a similar manner as it was done
for uniform networks. 
Another possible extension of the MBN algorithm is to choose the order of edge addition according to certain rules that facilitate the formation of certain
structures, instead of randomly picking the node to be updated. This was done in a similar framework as the present work in \cite{maki-marttunen2013structure}
to encourage the formation of loops of a certain length, a task that would require an unbearable amount of computation without such facilitation. Both of
these aspects are left for future work.

\section{Acknowledgements}
I thank Keijo Ruohonen for his comments on the work. 
TCSC resources were used for heavy computations.
Part of the work was done while receiving funding from TISE graduate school and KAUTE foundation.

\bibliography{maki-marttunen2016algorithm}
\bibliographystyle{unsrt}

\newpage

\renewcommand{\theequation}{S\arabic{equation}}
\renewcommand{\thefigure}{S\arabic{figure}}
\renewcommand{\thetable}{S\arabic{table}}
\setcounter{equation}{0}
\setcounter{figure}{0}
\setcounter{table}{0}

\section{Supplemental material}

\subsection{Calculation of the score for possible input nodes: Three-node motifs}

\label{sec_matrix_G}
As shown in the function Calculatepoints, the points for a possible input node $i$ are calculated as
\begin{equation}\lambda_i =\sum_{r=1}^{N_{\mathrm{premot}}}\sum_{m=1}^{N_{\mathrm{mot}}}Q_{ir}G_{rm}w_m\label{eq_lambda}.\end{equation}
This equation consists of three entities: The weighting vector, $w\in\R^{N_{\mathrm{mot}}}$ (constant throughout the algorithm),
the dependency matrix between pre-motifs and motifs, $G\in\R^{N_{\mathrm{premot}}\times N_{\mathrm{mot}}}$ (constant throughout the algorithm),
and the matrix of numbers of each existing pre-motif, $Q\in\R^{N\times N_{\mathrm{premot}}}$ (changes every iteration).
Table \ref{AB3} lists all three-node pre-motifs ($N_{\mathrm{premot}}=32$) and motifs ($N_{\mathrm{mot}}=16$) and shows which
motifs would be formed ('+') or abolished ('--') if an edge was drawn from $i$ to $k$ in each of the pre-motifs.
The matrix $G$ corresponding to Table \ref{AB3} is as follows:\\
{\begin{center}
$G = $
\begin{tiny}
\begin{tabular}{|cccccccccccccccc|}
-1 \ha  1 \ha  0 \ha  0 \ha  0 \ha  0 \ha  0 \ha  0 \ha  0 \ha  0 \ha  0 \ha  0 \ha  0 \ha  0 \ha  0 \ha  0 \\
 0 \ha -1 \ha  0 \ha  0 \ha  1 \ha  0 \ha  0 \ha  0 \ha  0 \ha  0 \ha  0 \ha  0 \ha  0 \ha  0 \ha  0 \ha  0 \\
 0 \ha -1 \ha  0 \ha  0 \ha  0 \ha  1 \ha  0 \ha  0 \ha  0 \ha  0 \ha  0 \ha  0 \ha  0 \ha  0 \ha  0 \ha  0 \\
 0 \ha  0 \ha  0 \ha -1 \ha  0 \ha  0 \ha  0 \ha  0 \ha  1 \ha  0 \ha  0 \ha  0 \ha  0 \ha  0 \ha  0 \ha  0 \\
 0 \ha -1 \ha  0 \ha  0 \ha  1 \ha  0 \ha  0 \ha  0 \ha  0 \ha  0 \ha  0 \ha  0 \ha  0 \ha  0 \ha  0 \ha  0 \\
 0 \ha  0 \ha  0 \ha  0 \ha -1 \ha  0 \ha  0 \ha  0 \ha  0 \ha  1 \ha  0 \ha  0 \ha  0 \ha  0 \ha  0 \ha  0 \\
 0 \ha  0 \ha -1 \ha  0 \ha  0 \ha  0 \ha  0 \ha  1 \ha  0 \ha  0 \ha  0 \ha  0 \ha  0 \ha  0 \ha  0 \ha  0 \\
 0 \ha  0 \ha  0 \ha  0 \ha  0 \ha  0 \ha -1 \ha  0 \ha  0 \ha  0 \ha  0 \ha  0 \ha  1 \ha  0 \ha  0 \ha  0 \\
 0 \ha -1 \ha  1 \ha  0 \ha  0 \ha  0 \ha  0 \ha  0 \ha  0 \ha  0 \ha  0 \ha  0 \ha  0 \ha  0 \ha  0 \ha  0 \\
 0 \ha  0 \ha  0 \ha  0 \ha  0 \ha -1 \ha  0 \ha  1 \ha  0 \ha  0 \ha  0 \ha  0 \ha  0 \ha  0 \ha  0 \ha  0 \\
 0 \ha  0 \ha  0 \ha  0 \ha -1 \ha  0 \ha  0 \ha  1 \ha  0 \ha  0 \ha  0 \ha  0 \ha  0 \ha  0 \ha  0 \ha  0 \\
 0 \ha  0 \ha  0 \ha  0 \ha  0 \ha  0 \ha  0 \ha  0 \ha -1 \ha  0 \ha  0 \ha  0 \ha  0 \ha  1 \ha  0 \ha  0 \\
 0 \ha  0 \ha  0 \ha -1 \ha  0 \ha  0 \ha  1 \ha  0 \ha  0 \ha  0 \ha  0 \ha  0 \ha  0 \ha  0 \ha  0 \ha  0 \\
 0 \ha  0 \ha  0 \ha  0 \ha  0 \ha  0 \ha  0 \ha  0 \ha -1 \ha  0 \ha  0 \ha  0 \ha  1 \ha  0 \ha  0 \ha  0 \\
 0 \ha  0 \ha  0 \ha  0 \ha  0 \ha  0 \ha -1 \ha  0 \ha  0 \ha  0 \ha  1 \ha  0 \ha  0 \ha  0 \ha  0 \ha  0 \\
 0 \ha  0 \ha  0 \ha  0 \ha  0 \ha  0 \ha  0 \ha  0 \ha  0 \ha  0 \ha  0 \ha -1 \ha  0 \ha  0 \ha  1 \ha  0 \\ %
 0 \ha -1 \ha  0 \ha  1 \ha  0 \ha  0 \ha  0 \ha  0 \ha  0 \ha  0 \ha  0 \ha  0 \ha  0 \ha  0 \ha  0 \ha  0 \\
 0 \ha  0 \ha -1 \ha  0 \ha  0 \ha  0 \ha  1 \ha  0 \ha  0 \ha  0 \ha  0 \ha  0 \ha  0 \ha  0 \ha  0 \ha  0 \\
 0 \ha  0 \ha  0 \ha  0 \ha -1 \ha  0 \ha  0 \ha  0 \ha  1 \ha  0 \ha  0 \ha  0 \ha  0 \ha  0 \ha  0 \ha  0 \\
 0 \ha  0 \ha  0 \ha  0 \ha  0 \ha  0 \ha -1 \ha  0 \ha  0 \ha  0 \ha  0 \ha  1 \ha  0 \ha  0 \ha  0 \ha  0 \\
 0 \ha  0 \ha  0 \ha  0 \ha  0 \ha -1 \ha  0 \ha  0 \ha  1 \ha  0 \ha  0 \ha  0 \ha  0 \ha  0 \ha  0 \ha  0 \\
 0 \ha  0 \ha  0 \ha  0 \ha  0 \ha  0 \ha  0 \ha -1 \ha  0 \ha  0 \ha  0 \ha  0 \ha  1 \ha  0 \ha  0 \ha  0 \\
 0 \ha  0 \ha  0 \ha  0 \ha  0 \ha  0 \ha  0 \ha -1 \ha  0 \ha  0 \ha  0 \ha  0 \ha  0 \ha  1 \ha  0 \ha  0 \\
 0 \ha  0 \ha  0 \ha  0 \ha  0 \ha  0 \ha  0 \ha  0 \ha  0 \ha  0 \ha -1 \ha  0 \ha  0 \ha  0 \ha  1 \ha  0 \\
 0 \ha  0 \ha  0 \ha  0 \ha -1 \ha  0 \ha  1 \ha  0 \ha  0 \ha  0 \ha  0 \ha  0 \ha  0 \ha  0 \ha  0 \ha  0 \\
 0 \ha  0 \ha  0 \ha  0 \ha  0 \ha  0 \ha  0 \ha -1 \ha  0 \ha  0 \ha  1 \ha  0 \ha  0 \ha  0 \ha  0 \ha  0 \\
 0 \ha  0 \ha  0 \ha  0 \ha  0 \ha  0 \ha  0 \ha  0 \ha  0 \ha -1 \ha  0 \ha  0 \ha  1 \ha  0 \ha  0 \ha  0 \\
 0 \ha  0 \ha  0 \ha  0 \ha  0 \ha  0 \ha  0 \ha  0 \ha  0 \ha  0 \ha  0 \ha  0 \ha -1 \ha  0 \ha  1 \ha  0 \\
 0 \ha  0 \ha  0 \ha  0 \ha  0 \ha  0 \ha  0 \ha  0 \ha -1 \ha  0 \ha  0 \ha  1 \ha  0 \ha  0 \ha  0 \ha  0 \\
 0 \ha  0 \ha  0 \ha  0 \ha  0 \ha  0 \ha  0 \ha  0 \ha  0 \ha  0 \ha  0 \ha  0 \ha  0 \ha -1 \ha  1 \ha  0 \\
 0 \ha  0 \ha  0 \ha  0 \ha  0 \ha  0 \ha  0 \ha  0 \ha  0 \ha  0 \ha  0 \ha  0 \ha -1 \ha  0 \ha  1 \ha  0 \\
 0 \ha  0 \ha  0 \ha  0 \ha  0 \ha  0 \ha  0 \ha  0 \ha  0 \ha  0 \ha  0 \ha  0 \ha  0 \ha  0 \ha -1 \ha  1
\end{tabular}\end{tiny}.\end{center}

\begin{table}[!hb]
\pagenumbering{gobble}
\caption{Table showing whether a certain three-node motif would be formed (+) or destroyed (--) by drawing a connection from $i$ to $k$ in each case of a previously existing motif.}
\hspace{6pt}PRE-\hspace{160pt}MOTIF\\
MOTIF
\vspace{10pt}\\
\begin{tabular}{c|ccccccccccccccccc}
\vspace{-0pt}&\includegraphics[clip,trim=200pt 283pt 178pt 266pt, width=0.02\textwidth]{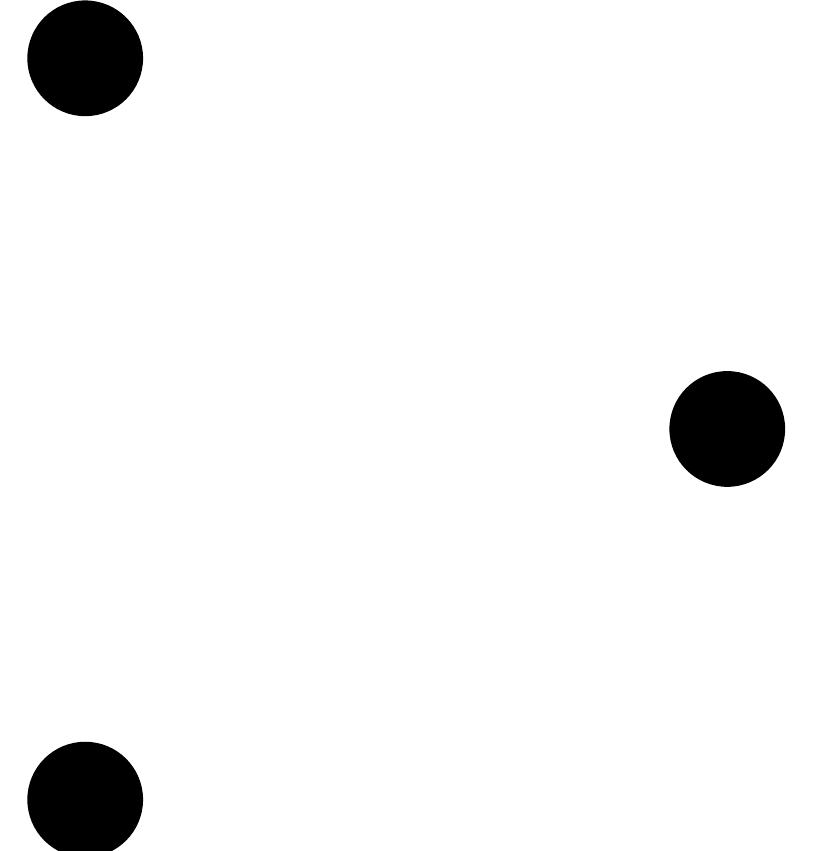}\hspace{-4pt}
\vspace{-0pt}&\hspace{-8pt}\includegraphics[clip,trim=200pt 283pt 178pt 266pt, width=0.02\textwidth]{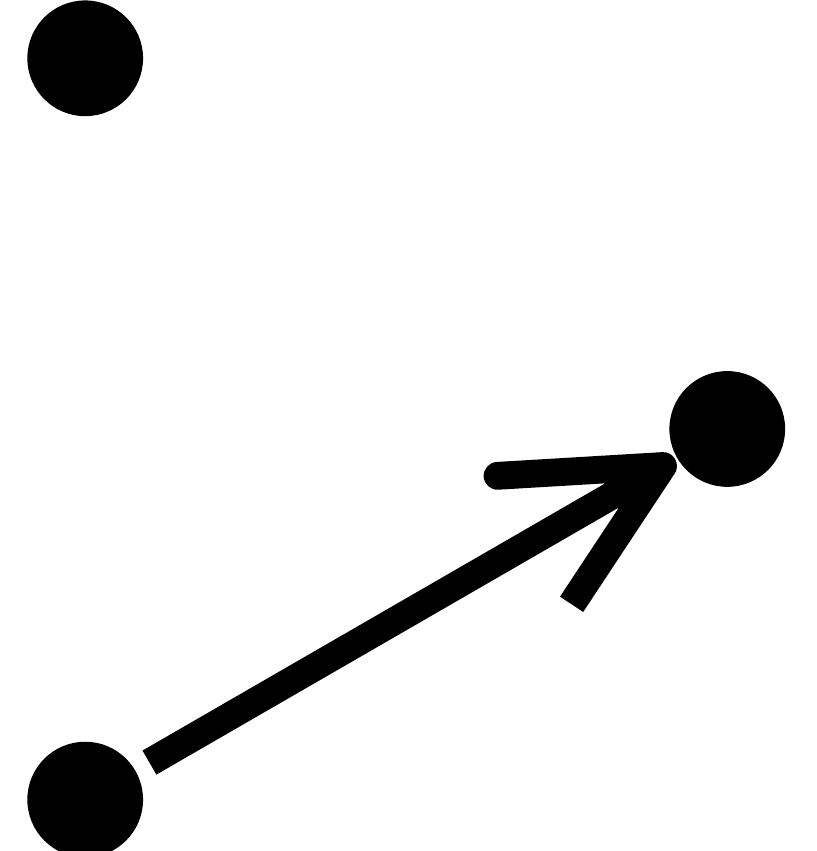}\hspace{-4pt}
\vspace{-0pt}&\hspace{-8pt}\includegraphics[clip,trim=200pt 283pt 178pt 266pt, width=0.02\textwidth]{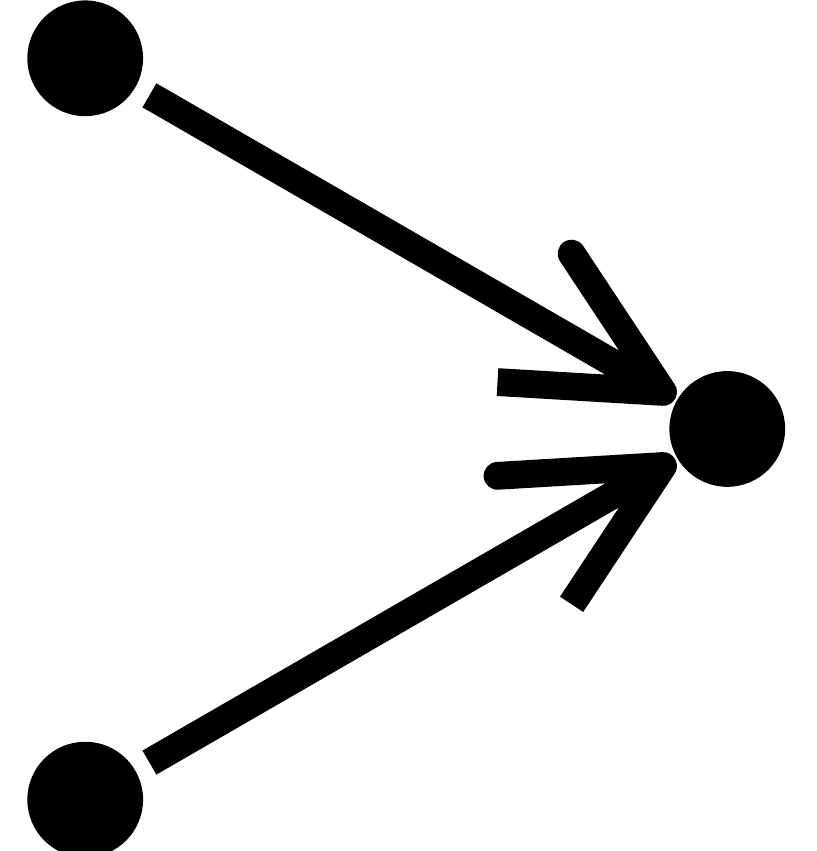}\hspace{-4pt}
\vspace{-0pt}&\hspace{-8pt}\includegraphics[clip,trim=200pt 283pt 178pt 266pt, width=0.02\textwidth]{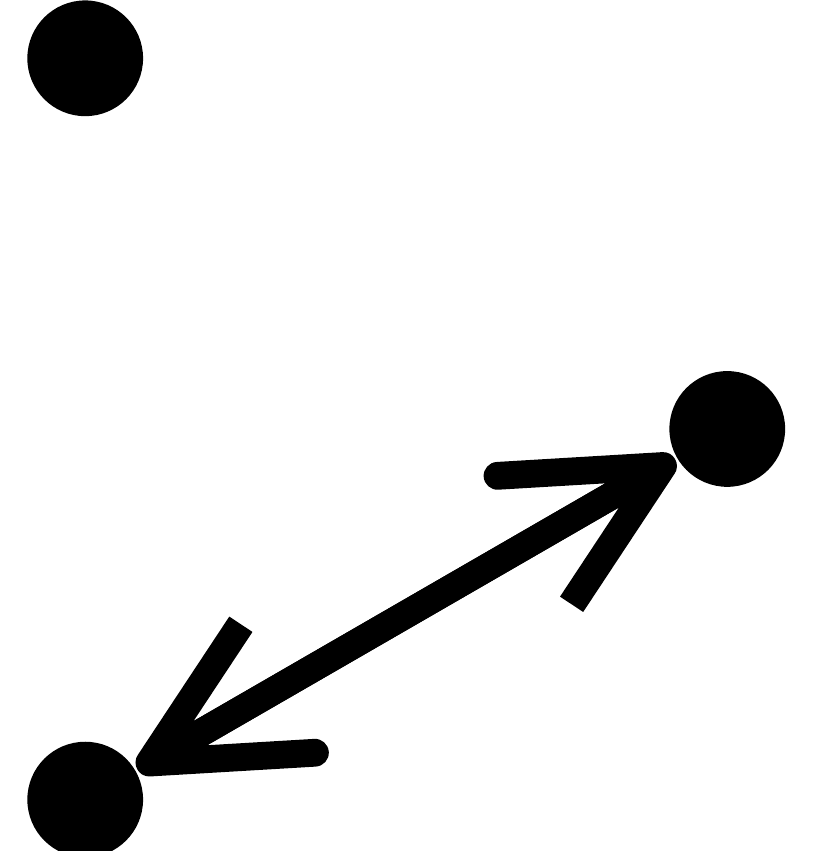}\hspace{-4pt}
\vspace{-0pt}&\hspace{-8pt}\includegraphics[clip,trim=200pt 283pt 178pt 266pt, width=0.02\textwidth]{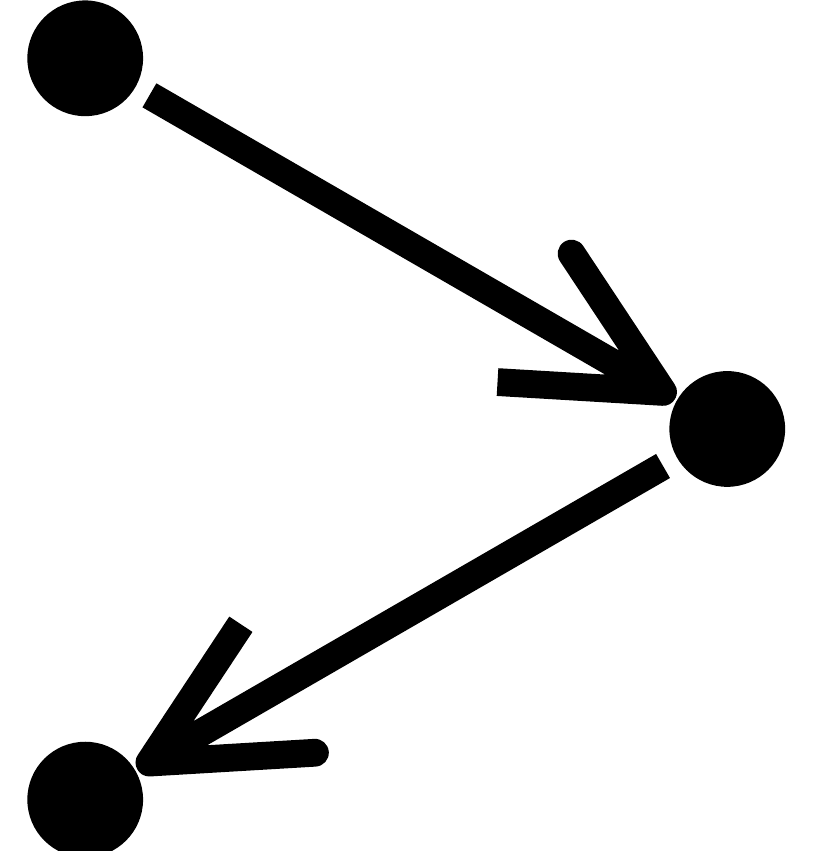}\hspace{-4pt}
\vspace{-0pt}&\hspace{-8pt}\includegraphics[clip,trim=200pt 283pt 178pt 266pt, width=0.02\textwidth]{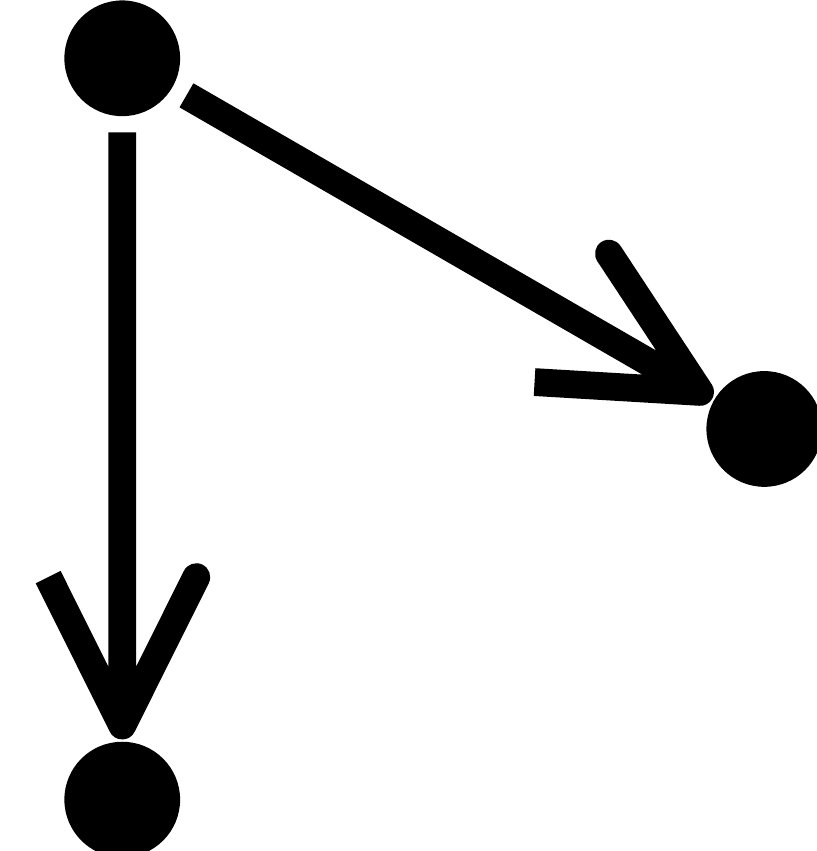}\hspace{-4pt}
\vspace{-0pt}&\hspace{-8pt}\includegraphics[clip,trim=200pt 283pt 178pt 266pt, width=0.02\textwidth]{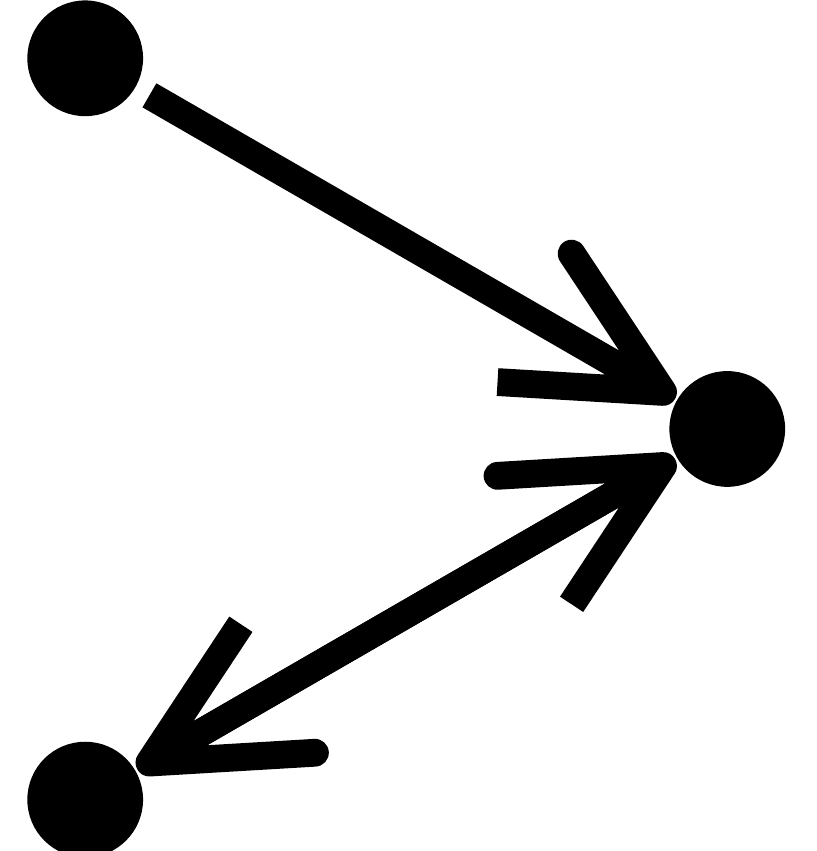}\hspace{-4pt}
\vspace{-0pt}&\hspace{-8pt}\includegraphics[clip,trim=200pt 283pt 178pt 266pt, width=0.02\textwidth]{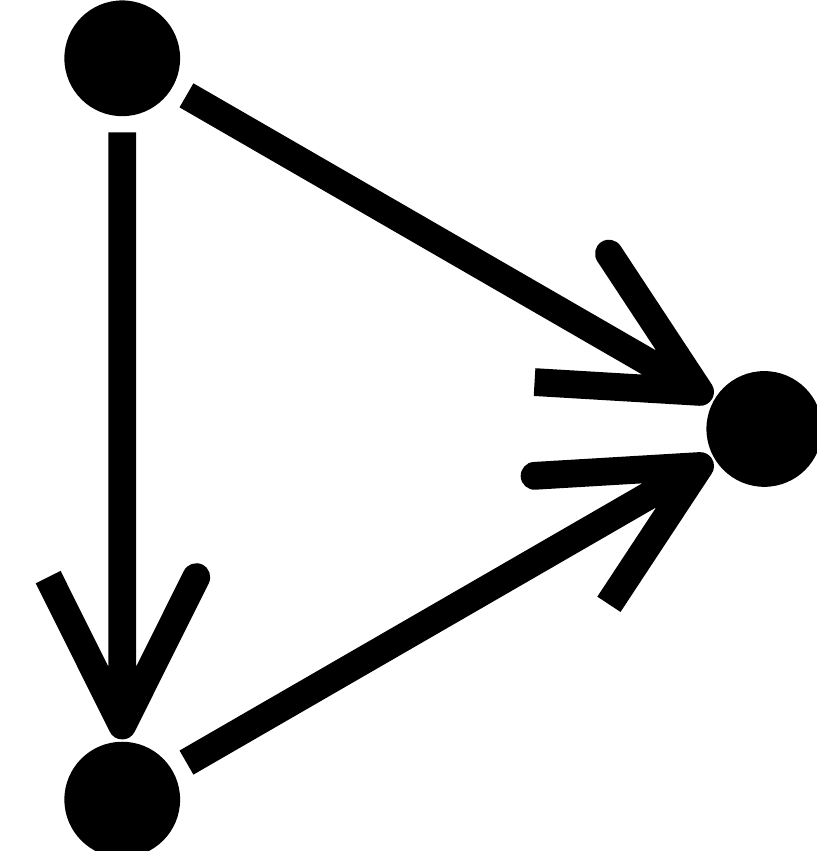}\hspace{-4pt}
\vspace{-0pt}&\hspace{-8pt}\includegraphics[clip,trim=200pt 283pt 178pt 266pt, width=0.02\textwidth]{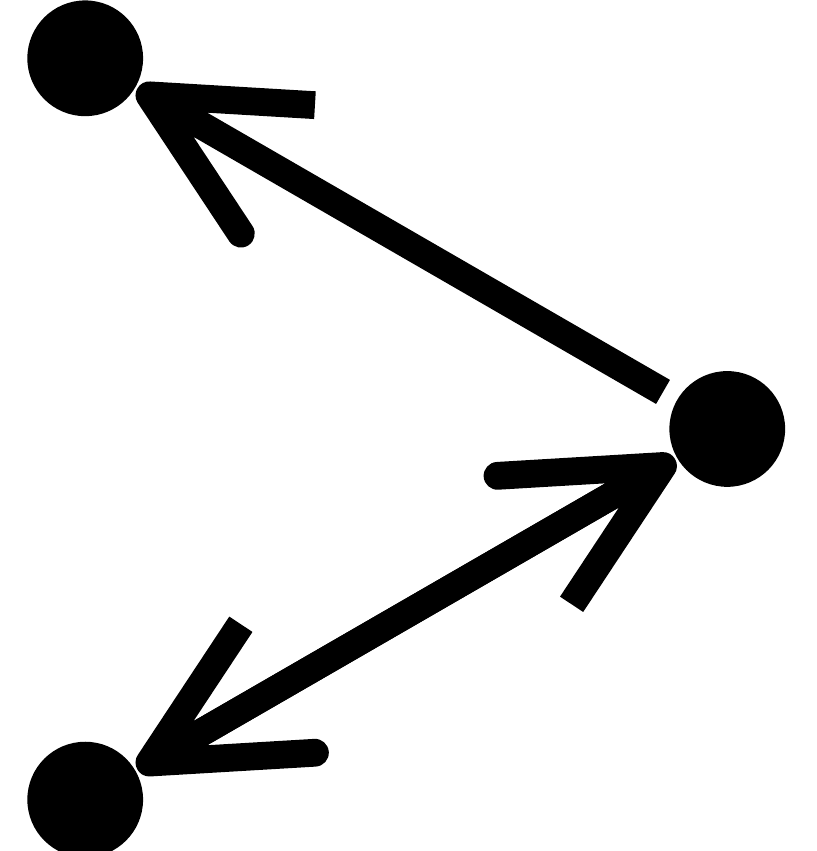}\hspace{-4pt}
\vspace{-0pt}&\hspace{-8pt}\includegraphics[clip,trim=200pt 283pt 178pt 266pt, width=0.02\textwidth]{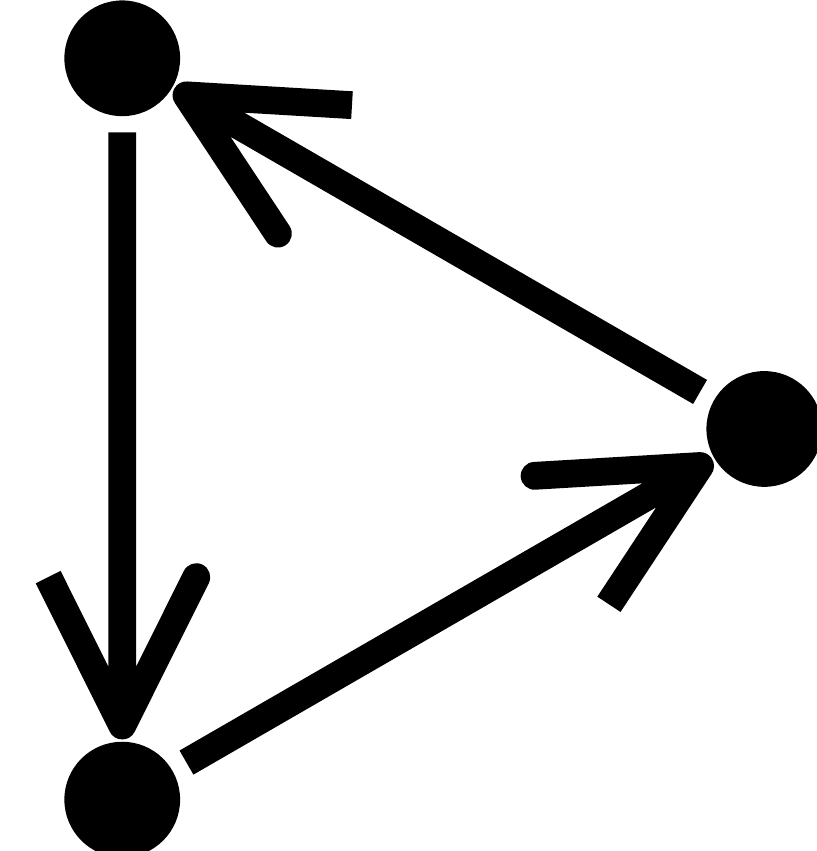}\hspace{-4pt}
\vspace{-0pt}&\hspace{-8pt}\includegraphics[clip,trim=200pt 283pt 178pt 266pt, width=0.02\textwidth]{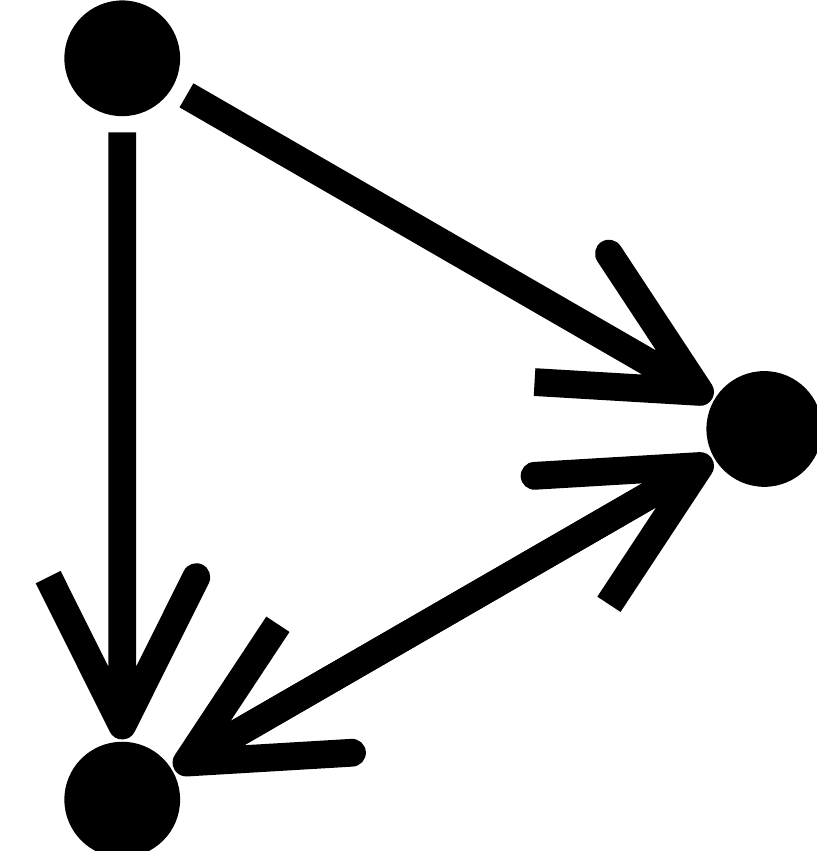}\hspace{-4pt}
\vspace{-0pt}&\hspace{-8pt}\includegraphics[clip,trim=200pt 283pt 178pt 266pt, width=0.02\textwidth]{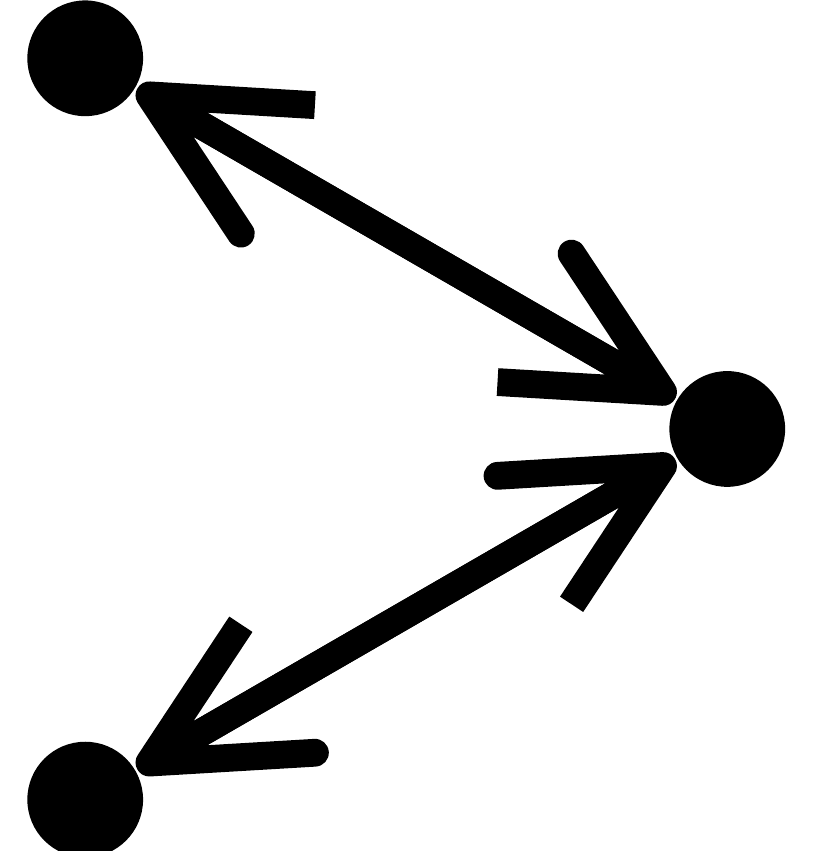}\hspace{-4pt}
\vspace{-0pt}&\hspace{-8pt}\includegraphics[clip,trim=200pt 283pt 178pt 266pt, width=0.02\textwidth]{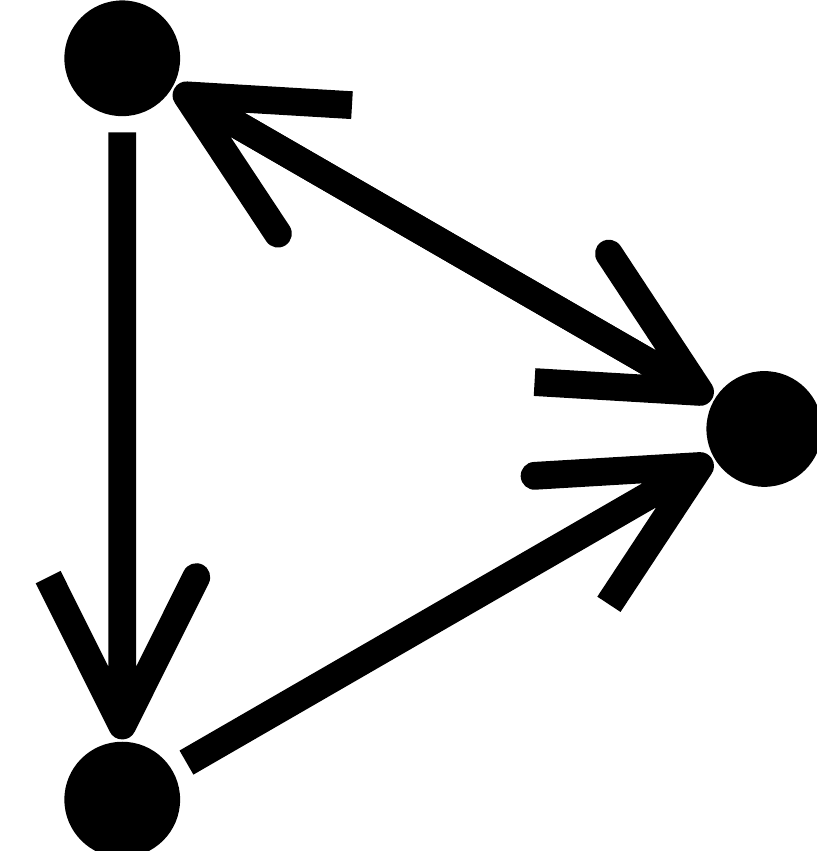}\hspace{-4pt}
\vspace{-0pt}&\hspace{-8pt}\includegraphics[clip,trim=200pt 283pt 178pt 266pt, width=0.02\textwidth]{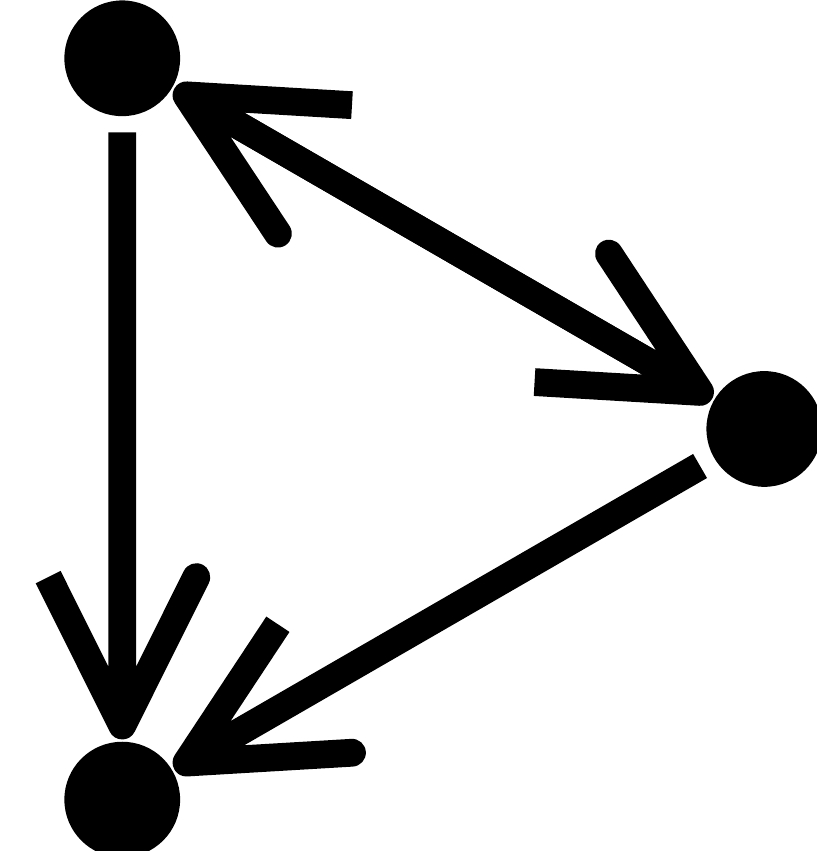}\hspace{-4pt}
\vspace{-0pt}&\hspace{-8pt}\includegraphics[clip,trim=200pt 283pt 178pt 266pt, width=0.02\textwidth]{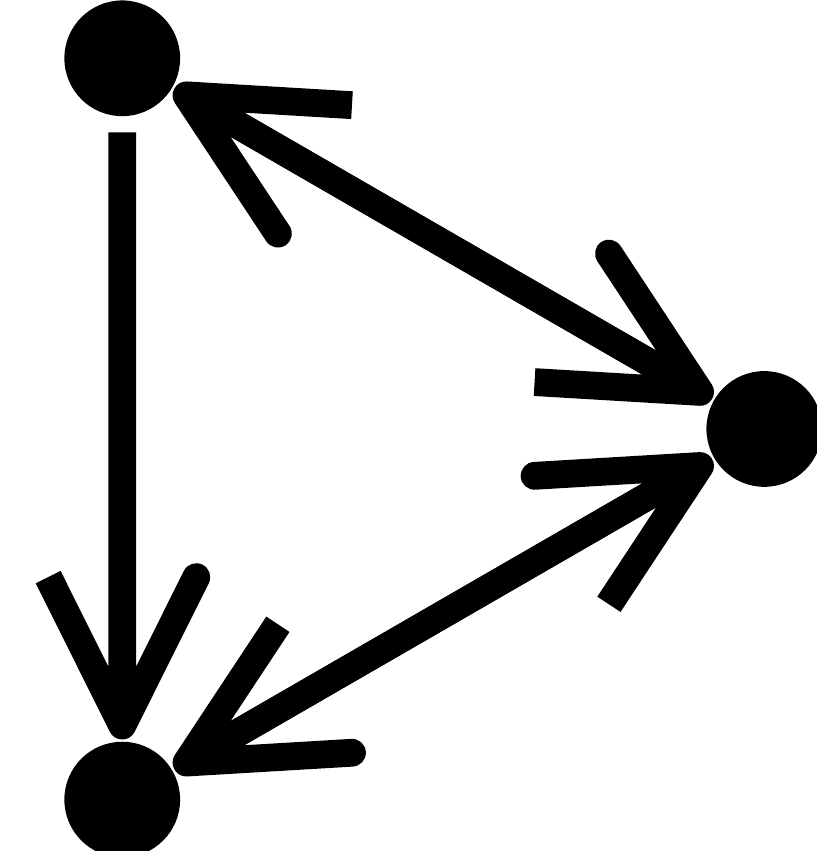}\hspace{-4pt}
\vspace{-0pt}&\hspace{-8pt}\includegraphics[clip,trim=200pt 283pt 178pt 266pt, width=0.02\textwidth]{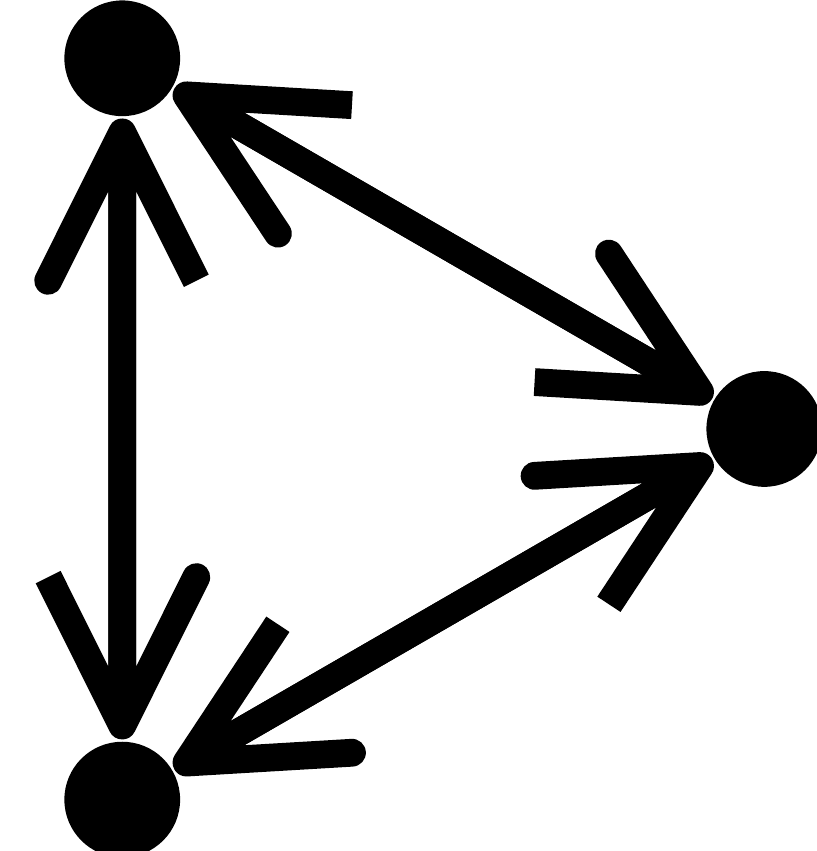}\hspace{-4pt}
\\
\hspace{-13pt}$i$ -- $j$ -- $k$ & \hspace{-6pt}\scriptsize 1\hspace{-6pt} & \hspace{-12pt}\scriptsize 2\hspace{-6pt} & \hspace{-12pt}\scriptsize3\hspace{-6pt} & \hspace{-12pt}\scriptsize4\hspace{-6pt} &
 \hspace{-12pt}\scriptsize5\hspace{-6pt} & \hspace{-12pt}\scriptsize6\hspace{-6pt} & \hspace{-12pt}\scriptsize7\hspace{-6pt} & \hspace{-12pt}\scriptsize8\hspace{-6pt} &
 \hspace{-12pt}\scriptsize9\hspace{-6pt} & \hspace{-12pt}\scriptsize{10}\hspace{-6pt} & \hspace{-12pt}\scriptsize{11}\hspace{-6pt} & \hspace{-12pt}\scriptsize{12}\hspace{-6pt} &
 \hspace{-12pt}\scriptsize{13}\hspace{-6pt} & \hspace{-12pt}\scriptsize{14}\hspace{-6pt} & \hspace{-12pt}\scriptsize{15}\hspace{-6pt} & \hspace{-6pt}\scriptsize{16}\\
\hline
\vspace{-1pt}\hspace{-13pt}\includegraphics[clip,trim=80pt 355pt 59pt 335pt, width=0.08\textwidth]{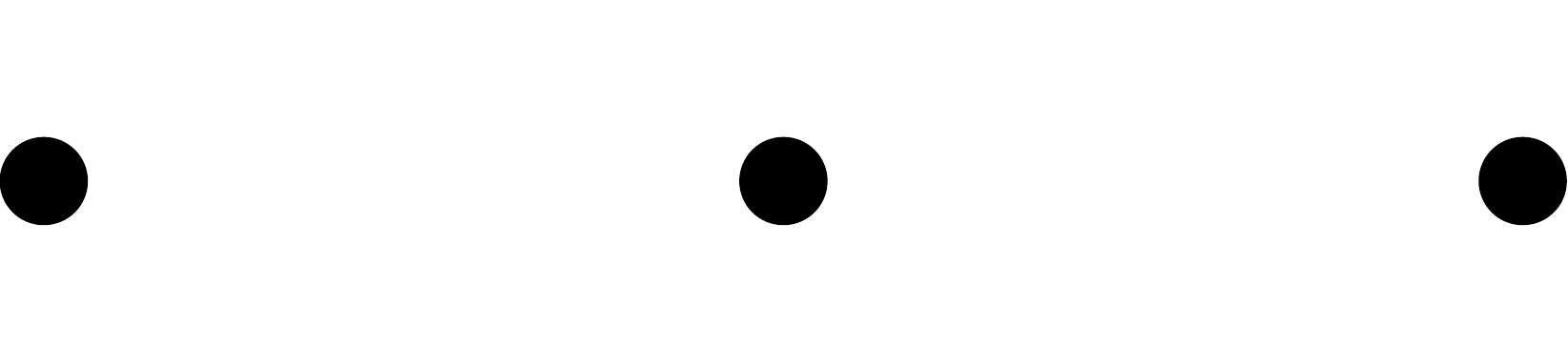}&\mm & \mp &   &   &   &   &   &   &   &   &   &   &   &   &   &  \\
\vspace{-1pt}\hspace{-13pt}\includegraphics[clip,trim=80pt 355pt 59pt 335pt, width=0.08\textwidth]{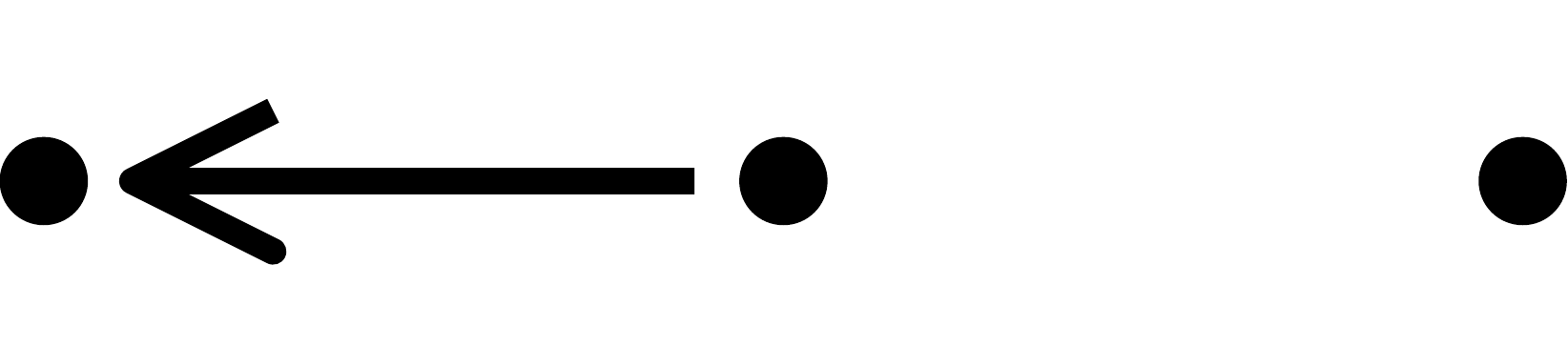}&   &\mm &   &   & \mp &   &   &   &   &   &   &   &   &   &   &  \\
\vspace{-1pt}\hspace{-13pt}\includegraphics[clip,trim=80pt 355pt 59pt 335pt, width=0.08\textwidth]{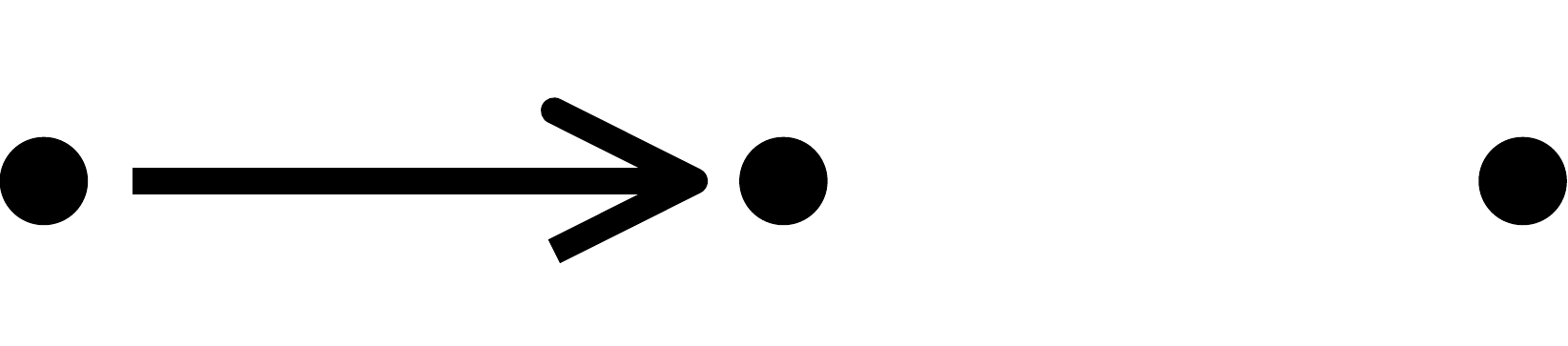}&   &\mm &   &   &   & \mp &   &   &   &   &   &   &   &   &   &  \\
\vspace{-1pt}\hspace{-13pt}\includegraphics[clip,trim=80pt 355pt 59pt 335pt, width=0.08\textwidth]{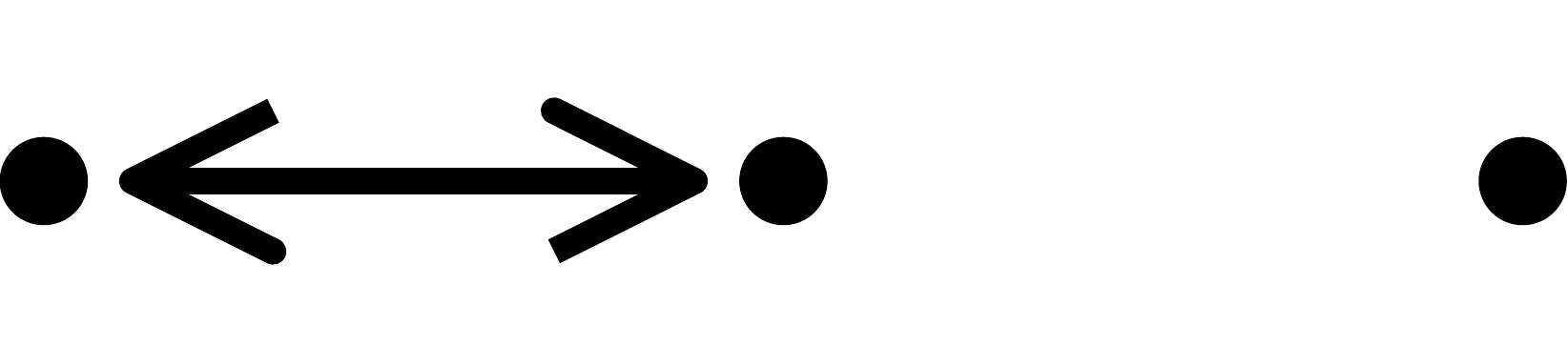}&   &   &   &\mm &   &   &   &   & \mp &   &   &   &   &   &   &  \\
\vspace{-1pt}\hspace{-13pt}\includegraphics[clip,trim=80pt 355pt 59pt 335pt, width=0.08\textwidth]{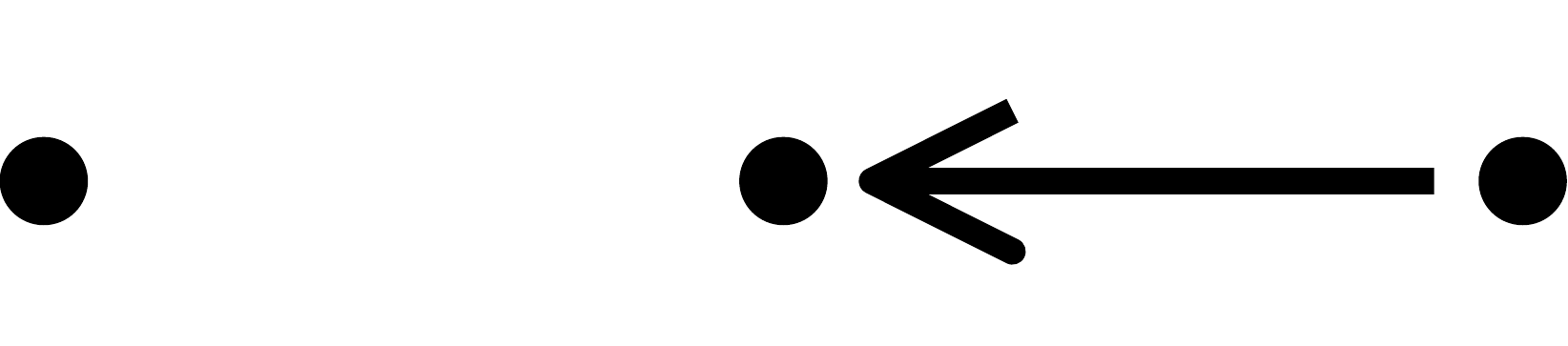}&   &\mm &   &   & \mp &   &   &   &   &   &   &   &   &   &   &  \\
\vspace{-1pt}\hspace{-13pt}\includegraphics[clip,trim=80pt 355pt 59pt 335pt, width=0.08\textwidth]{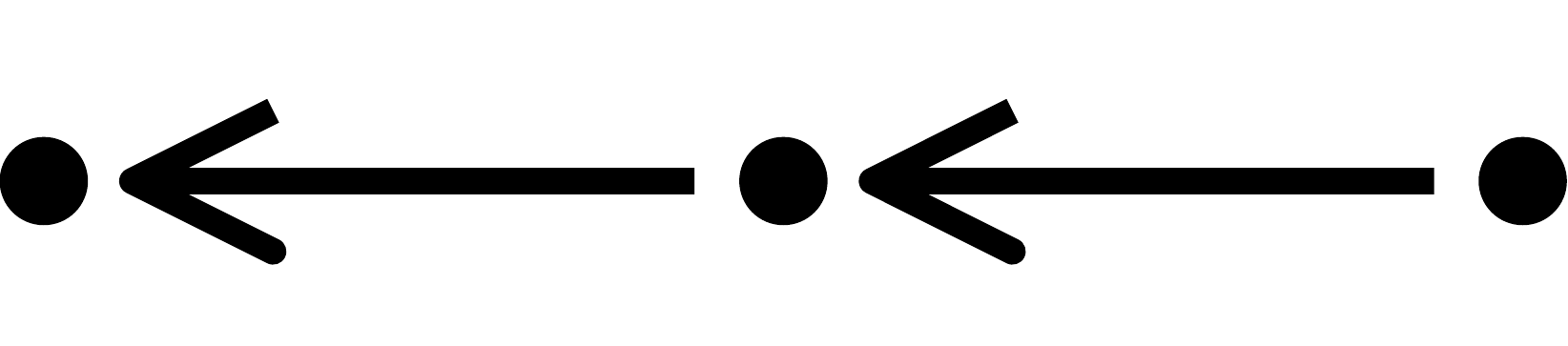}&   &   &   &   &\mm &   &   &   &   & \mp &   &   &   &   &   &  \\
\vspace{-1pt}\hspace{-13pt}\includegraphics[clip,trim=80pt 355pt 59pt 335pt, width=0.08\textwidth]{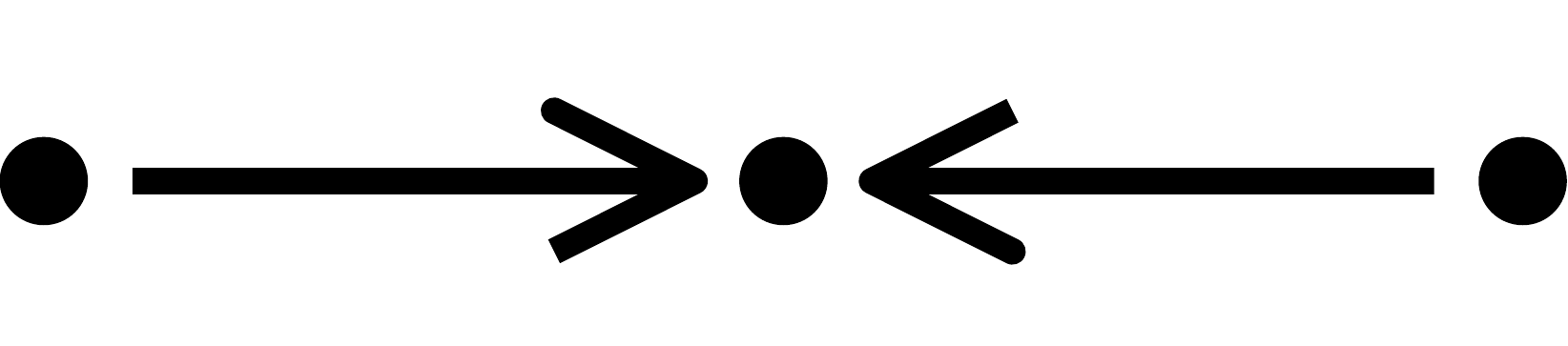}&   &   &\mm &   &   &   &   & \mp &   &   &   &   &   &   &   &  \\\
\vspace{-1pt}\hspace{-13pt}\includegraphics[clip,trim=80pt 355pt 59pt 335pt, width=0.08\textwidth]{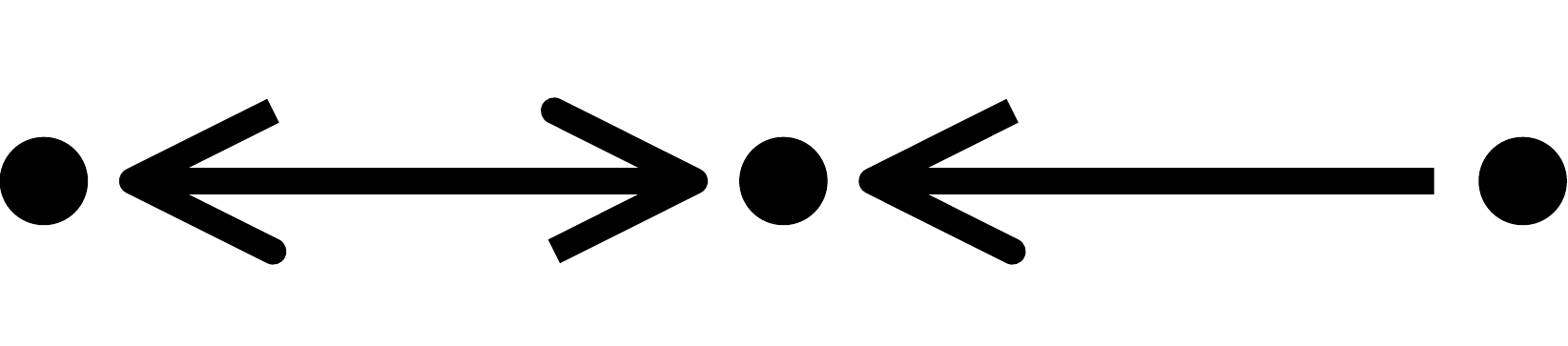}&   &   &   &   &   &   &\mm &   &   &   &   &   & \mp &   &   &  \\
\vspace{-1pt}\hspace{-13pt}\includegraphics[clip,trim=80pt 355pt 59pt 335pt, width=0.08\textwidth]{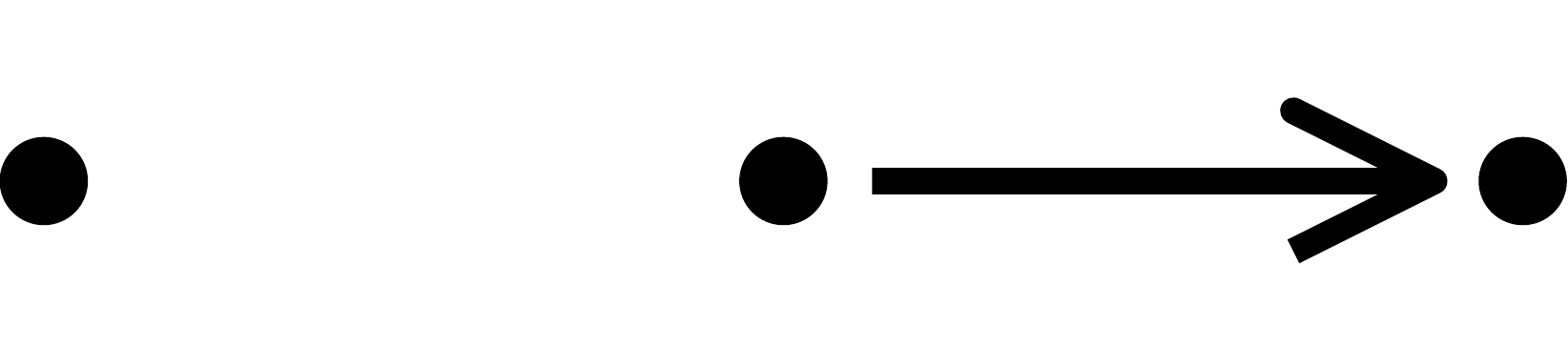}&   &\mm & \mp &   &   &   &   &   &   &   &   &   &   &   &   &  \\
\vspace{-1pt}\hspace{-13pt}\includegraphics[clip,trim=80pt 355pt 59pt 335pt, width=0.08\textwidth]{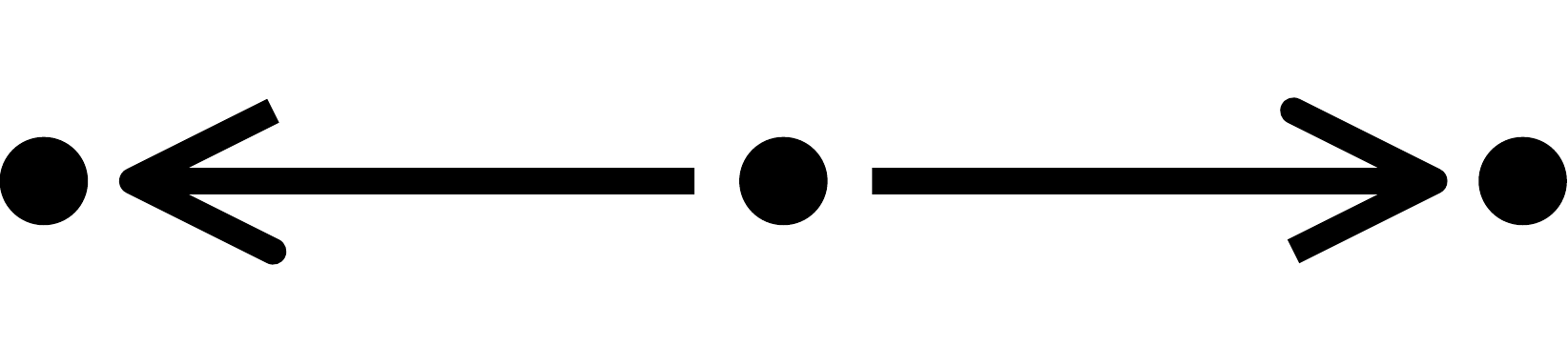}&   &   &   &   &   &\mm &   & \mp &   &   &   &   &   &   &   &  \\
\vspace{-1pt}\hspace{-13pt}\includegraphics[clip,trim=80pt 355pt 59pt 335pt, width=0.08\textwidth]{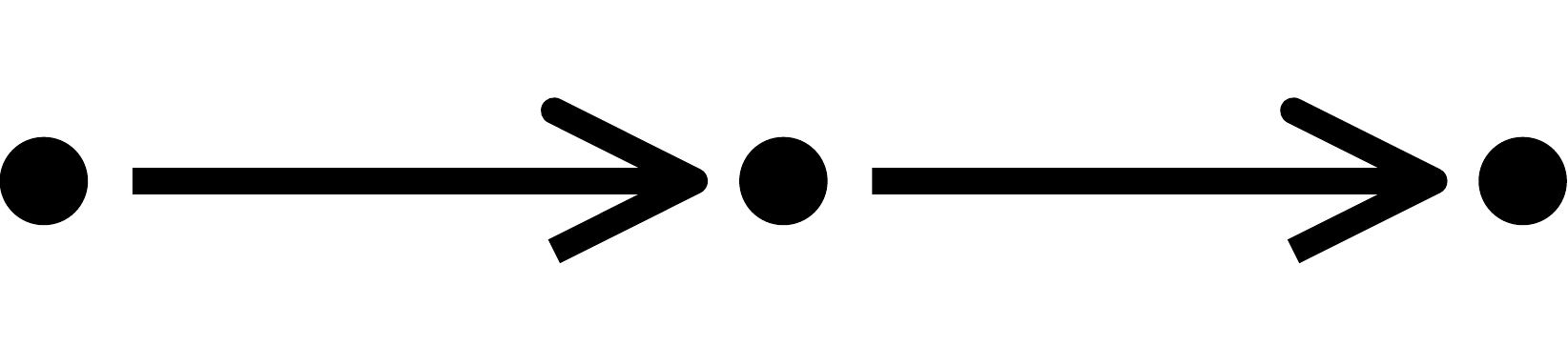}&   &   &   &   &\mm &   &   & \mp &   &   &   &   &   &   &   &  \\
\vspace{-1pt}\hspace{-13pt}\includegraphics[clip,trim=80pt 355pt 59pt 335pt, width=0.08\textwidth]{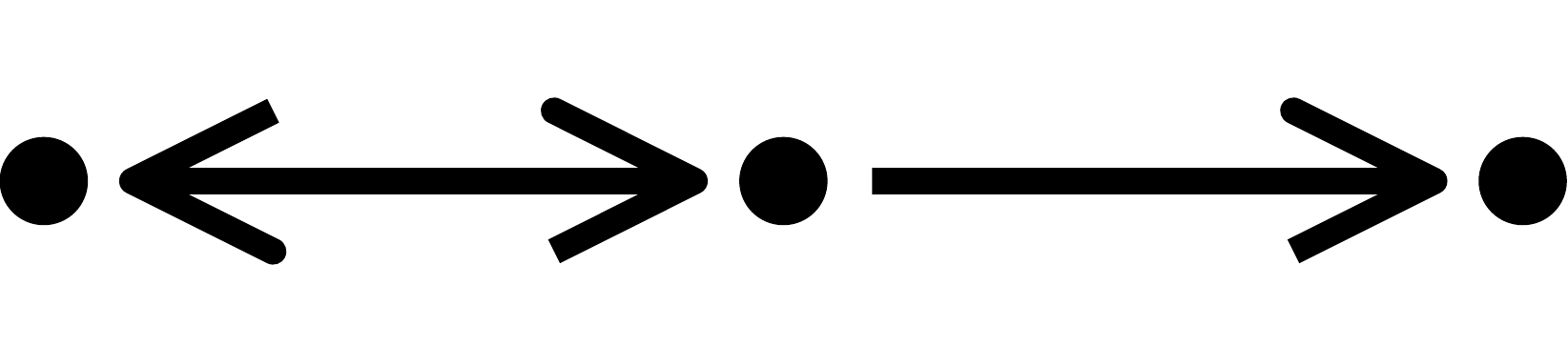}&   &   &   &   &   &   &   &   &\mm &   &   &   &   & \mp &   &  \\
\vspace{-1pt}\hspace{-13pt}\includegraphics[clip,trim=80pt 355pt 59pt 335pt, width=0.08\textwidth]{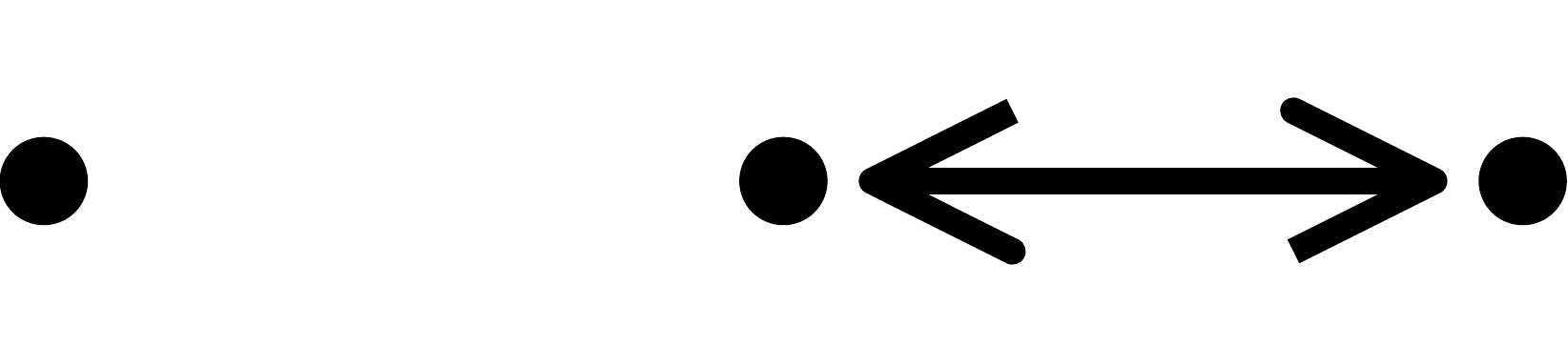}&   &   &   &\mm &   &   & \mp &   &   &   &   &   &   &   &   &  \\
\vspace{-1pt}\hspace{-13pt}\includegraphics[clip,trim=80pt 355pt 59pt 335pt, width=0.08\textwidth]{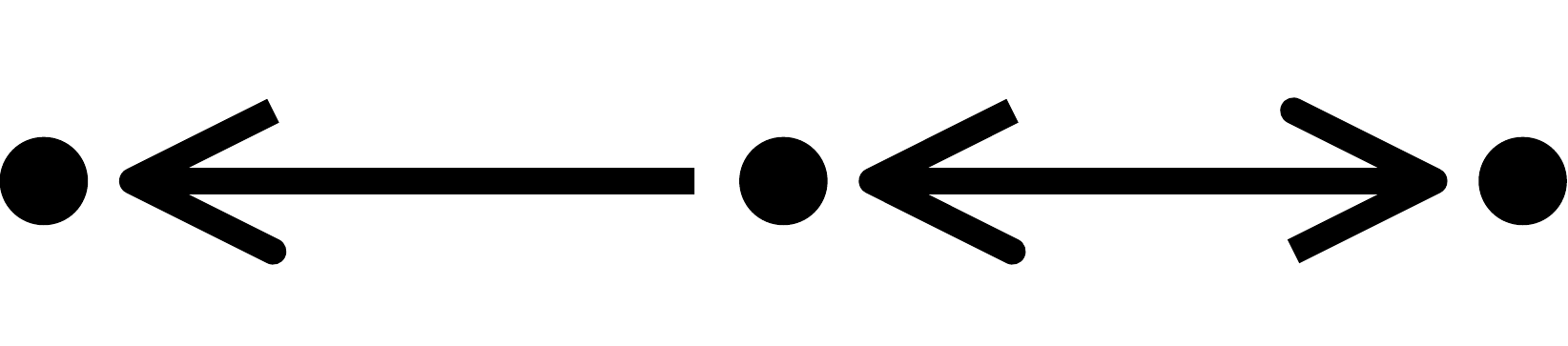}&   &   &   &   &   &   &   &   &\mm &   &   &   & \mp &   &   &  \\
\vspace{-1pt}\hspace{-13pt}\includegraphics[clip,trim=80pt 355pt 59pt 335pt, width=0.08\textwidth]{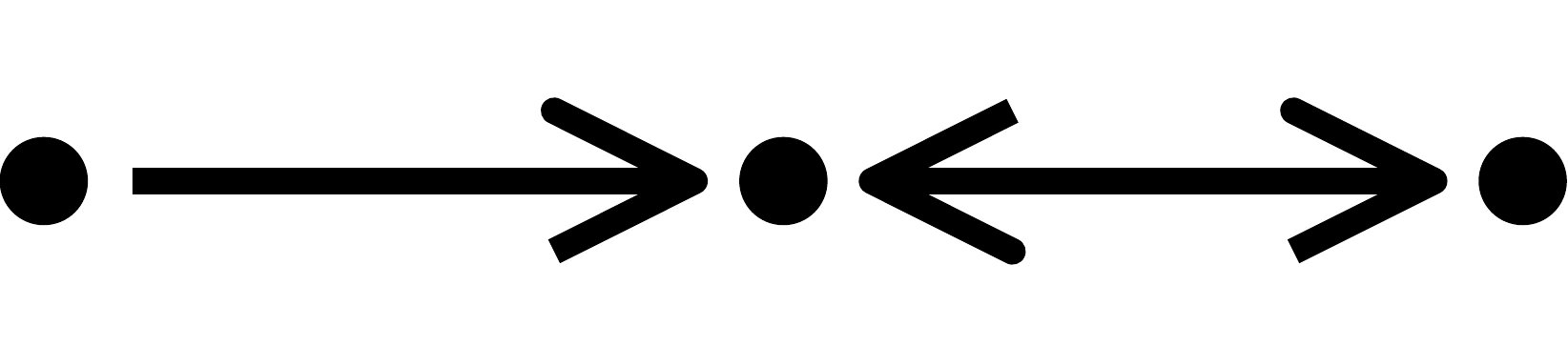}&   &   &   &   &   &   &\mm &   &   &   & \mp &   &   &   &   &  \\
\vspace{-1pt}\hspace{-13pt}\includegraphics[clip,trim=80pt 355pt 59pt 335pt, width=0.08\textwidth]{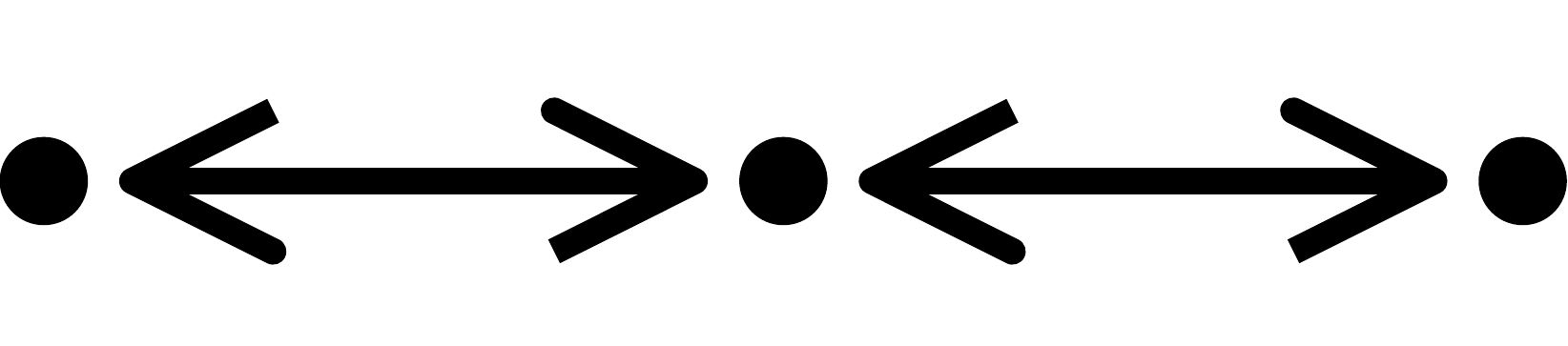}&   &   &   &   &   &   &   &   &   &   &   &\mm &   &   & \mp &  \\
\hline
\vspace{1pt}
\vspace{-1pt}\hspace{-13pt}\includegraphics[clip,trim=90pt 355pt 59pt 335pt, width=0.08\textwidth]{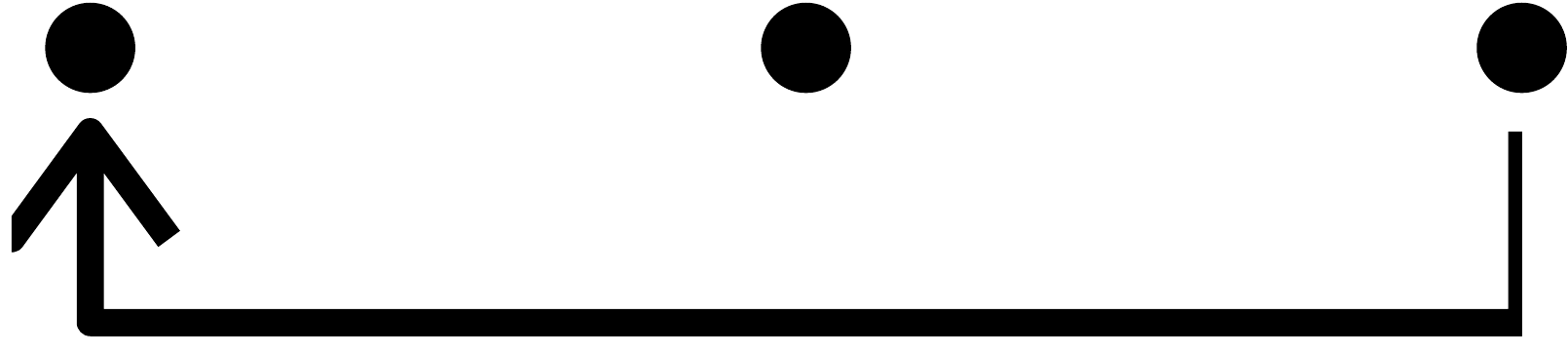}&   &\mm &   & \mp &   &   &   &   &   &   &   &   &   &   &   &  \\
\vspace{-1pt}\hspace{-13pt}\includegraphics[clip,trim=90pt 355pt 59pt 335pt, width=0.08\textwidth]{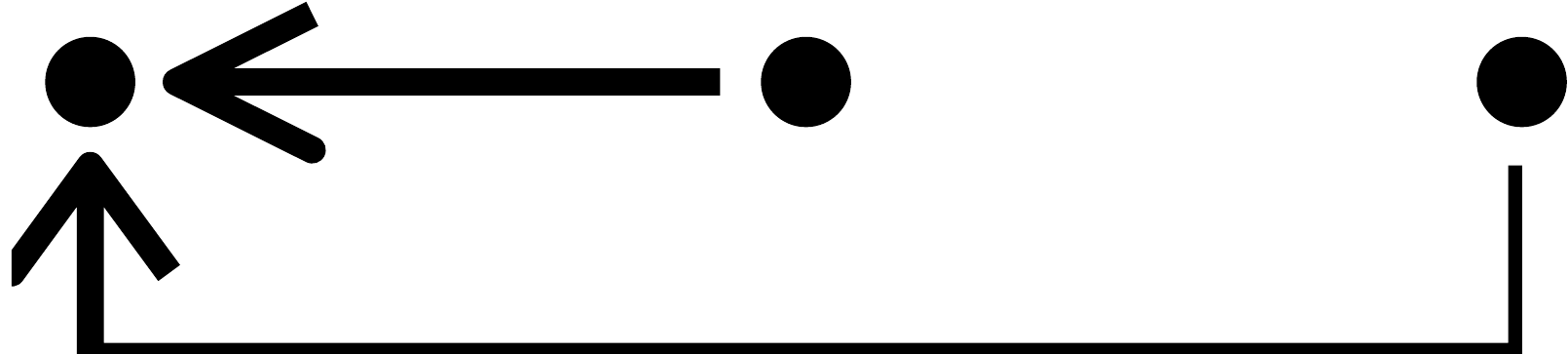}&   &   &\mm &   &   &   & \mp &   &   &   &   &   &   &   &   &  \\
\vspace{-1pt}\hspace{-13pt}\includegraphics[clip,trim=90pt 355pt 59pt 335pt, width=0.08\textwidth]{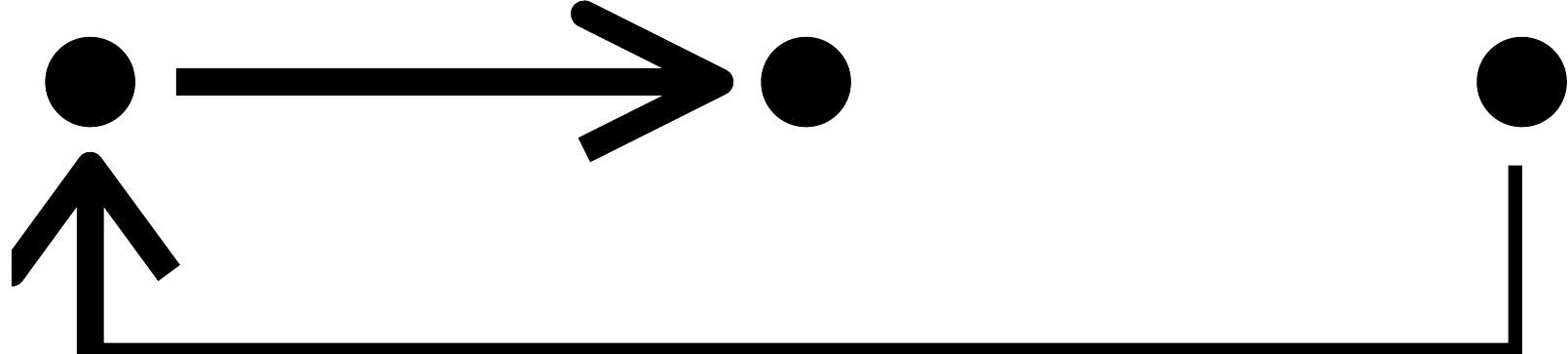}&   &   &   &   &\mm &   &   &   & \mp &   &   &   &   &   &   &  \\
\vspace{-1pt}\hspace{-13pt}\includegraphics[clip,trim=90pt 355pt 59pt 335pt, width=0.08\textwidth]{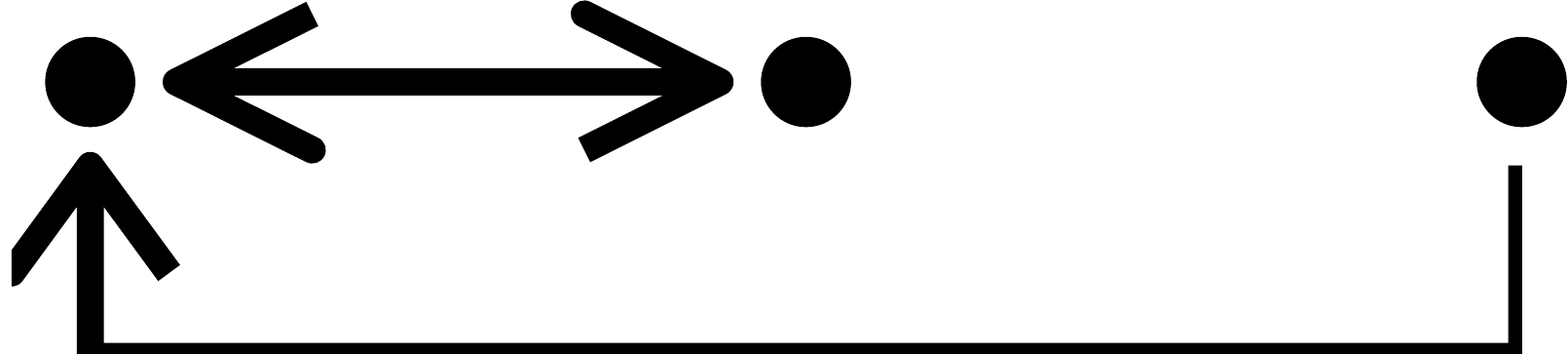}&   &   &   &   &   &   &\mm &   &   &   &   & \mp &   &   &   &  \\
\vspace{-1pt}\hspace{-13pt}\includegraphics[clip,trim=90pt 355pt 59pt 335pt, width=0.08\textwidth]{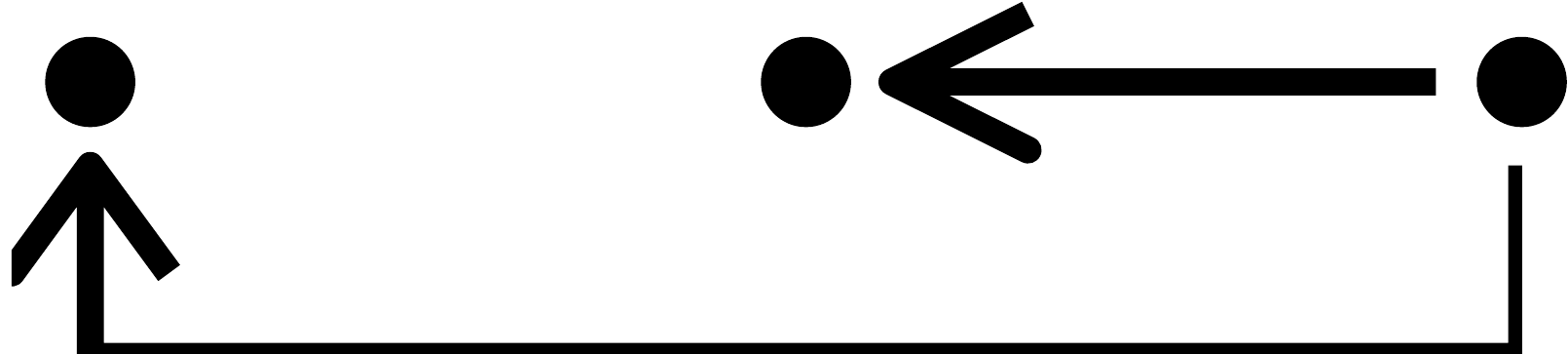}&   &   &   &   &   &\mm &   &   & \mp &   &   &   &   &   &   &  \\
\vspace{-1pt}\hspace{-13pt}\includegraphics[clip,trim=90pt 355pt 59pt 335pt, width=0.08\textwidth]{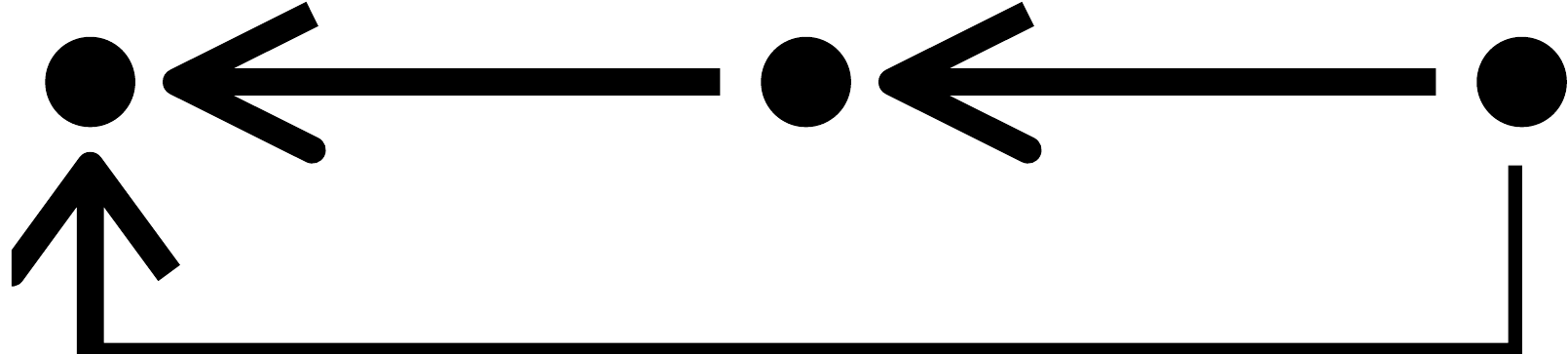}&   &   &   &   &   &   &   &\mm &   &   &   &   & \mp &   &   &  \\
\vspace{-1pt}\hspace{-13pt}\includegraphics[clip,trim=90pt 355pt 59pt 335pt, width=0.08\textwidth]{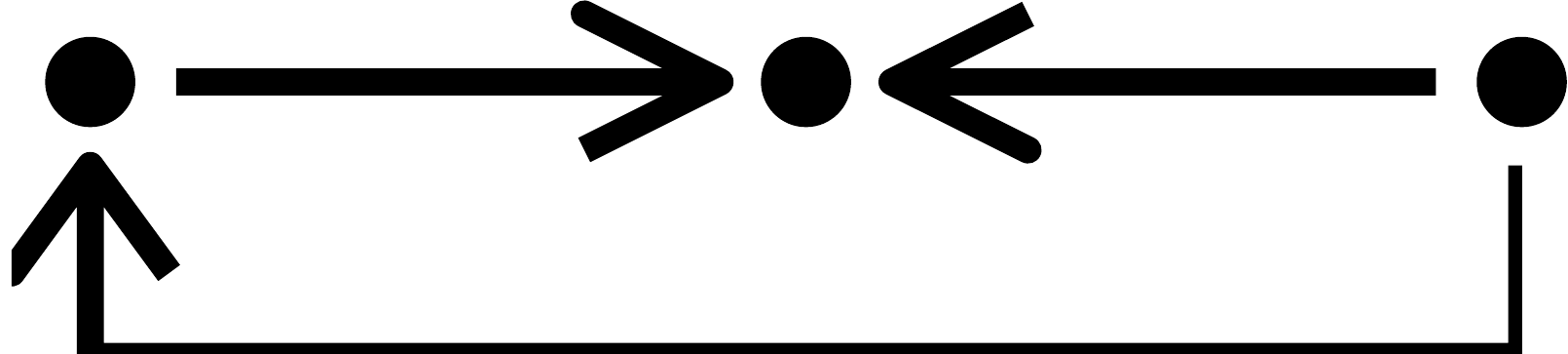}&   &   &   &   &   &   &   &\mm &   &   &   &   &   & \mp &   &  \\
\vspace{-1pt}\hspace{-13pt}\includegraphics[clip,trim=90pt 355pt 59pt 335pt, width=0.08\textwidth]{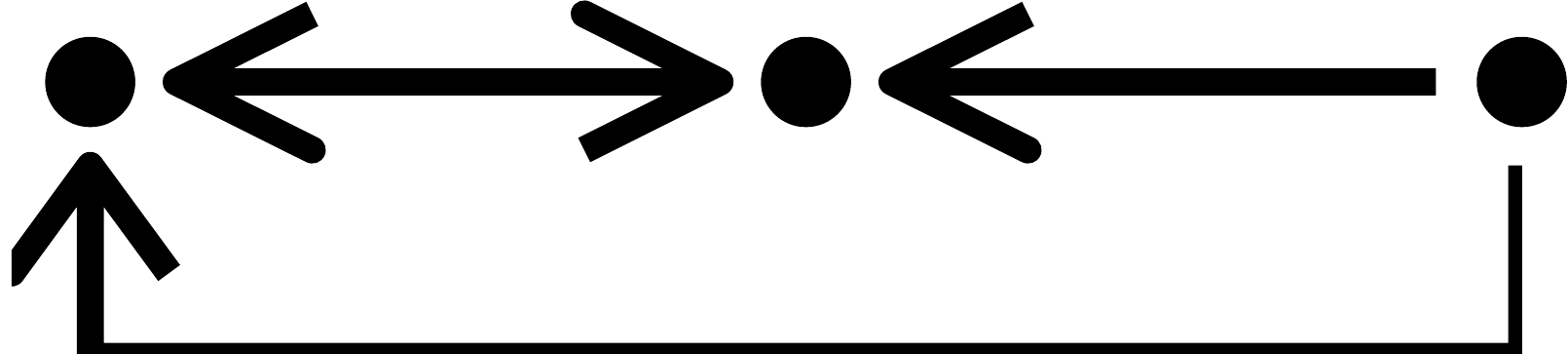}&   &   &   &   &   &   &   &   &   &   &\mm &   &   &   & \mp &  \\
\vspace{-1pt}\hspace{-13pt}\includegraphics[clip,trim=90pt 355pt 59pt 335pt, width=0.08\textwidth]{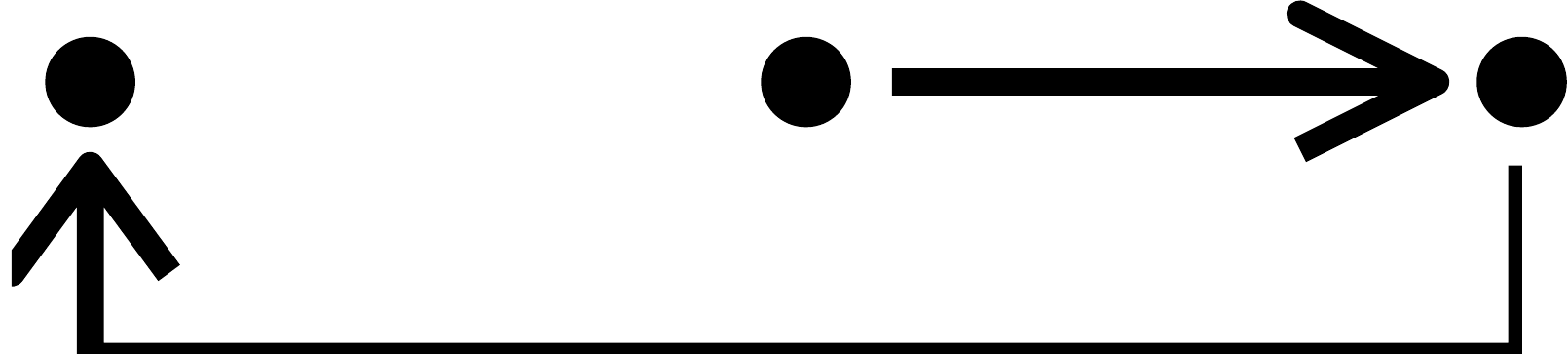}&   &   &   &   &\mm &   & \mp &   &   &   &   &   &   &   &   &  \\
\vspace{-1pt}\hspace{-13pt}\includegraphics[clip,trim=90pt 355pt 59pt 335pt, width=0.08\textwidth]{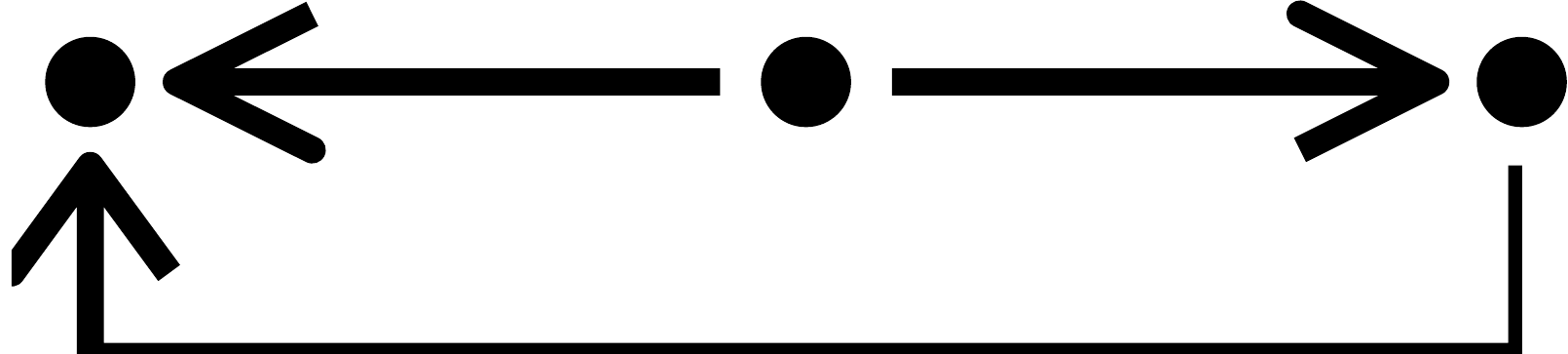}&   &   &   &   &   &   &   &\mm &   &   & \mp &   &   &   &   &  \\
\vspace{-1pt}\hspace{-13pt}\includegraphics[clip,trim=90pt 355pt 59pt 335pt, width=0.08\textwidth]{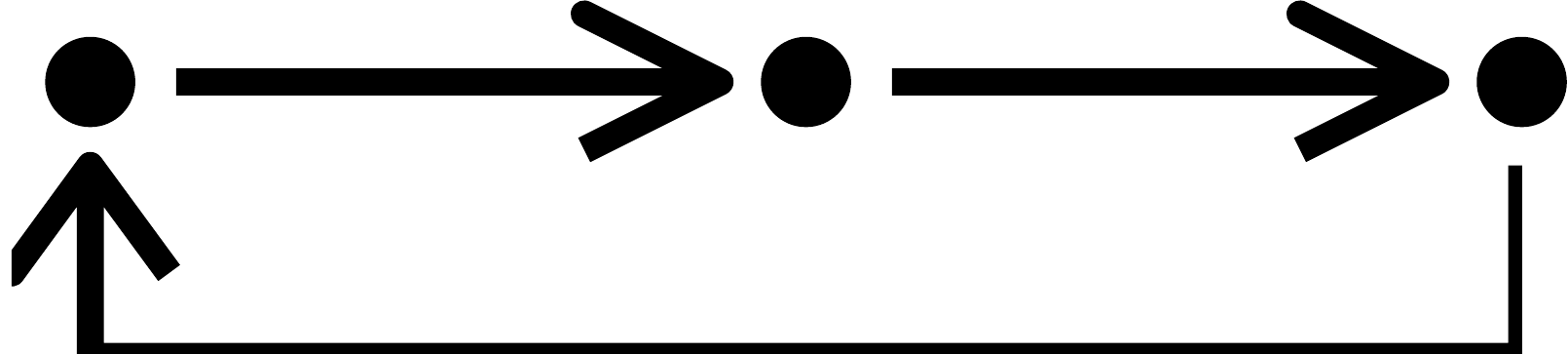}&   &   &   &   &   &   &   &   &   &\mm &   &   & \mp &   &   &  \\
\vspace{-1pt}\hspace{-13pt}\includegraphics[clip,trim=90pt 355pt 59pt 335pt, width=0.08\textwidth]{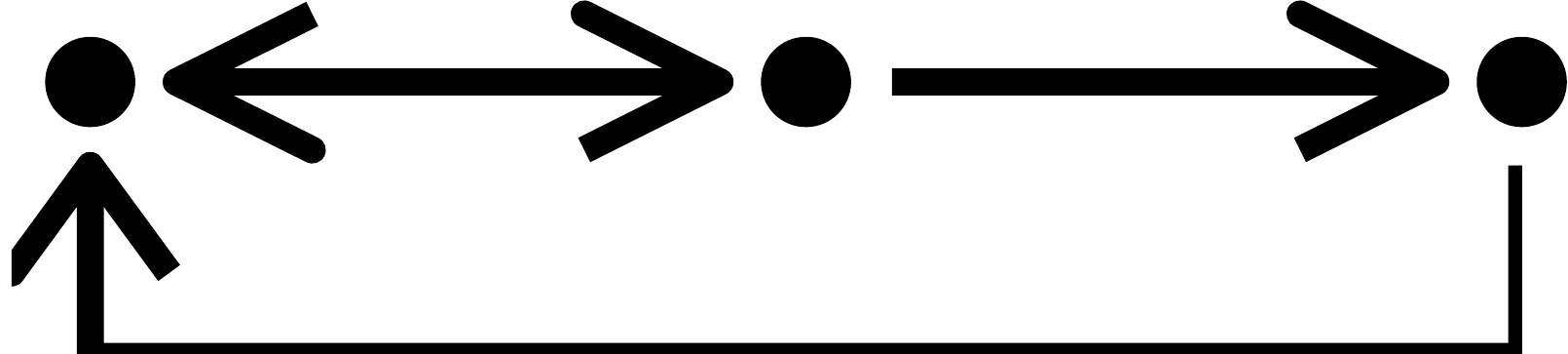}&   &   &   &   &   &   &   &   &   &   &   &   &\mm &   & \mp &  \\
\vspace{-1pt}\hspace{-13pt}\includegraphics[clip,trim=90pt 355pt 59pt 335pt, width=0.08\textwidth]{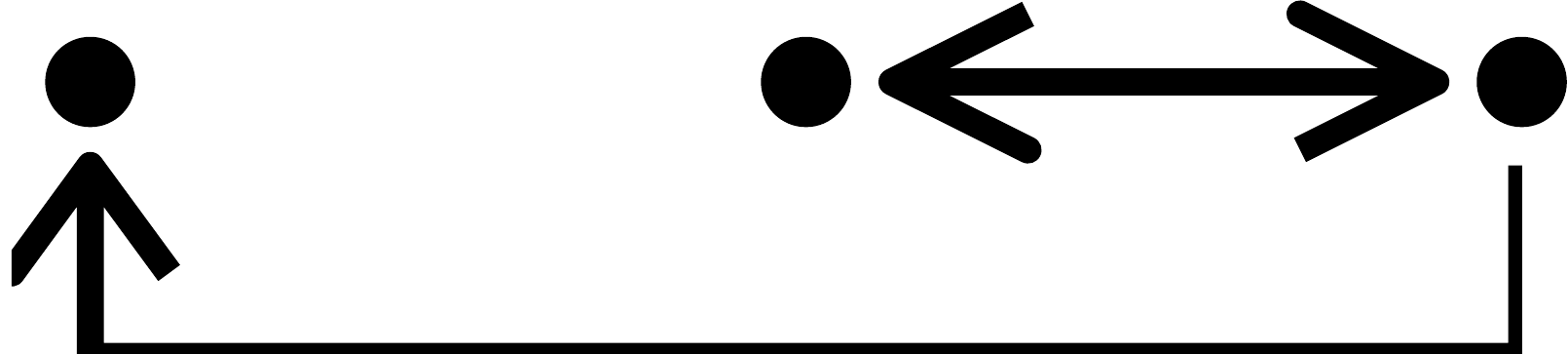}&   &   &   &   &   &   &   &   &\mm &   &   & \mp &   &   &   &  \\
\vspace{-1pt}\hspace{-13pt}\includegraphics[clip,trim=90pt 355pt 59pt 335pt, width=0.08\textwidth]{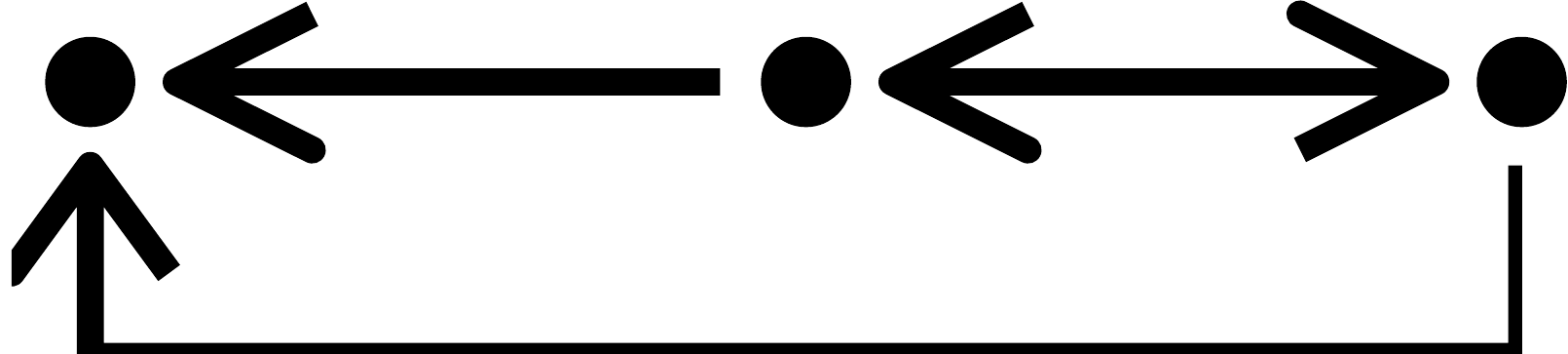}&   &   &   &   &   &   &   &   &   &   &   &   &   &\mm & \mp &  \\
\vspace{-1pt}\hspace{-13pt}\includegraphics[clip,trim=90pt 355pt 59pt 335pt, width=0.08\textwidth]{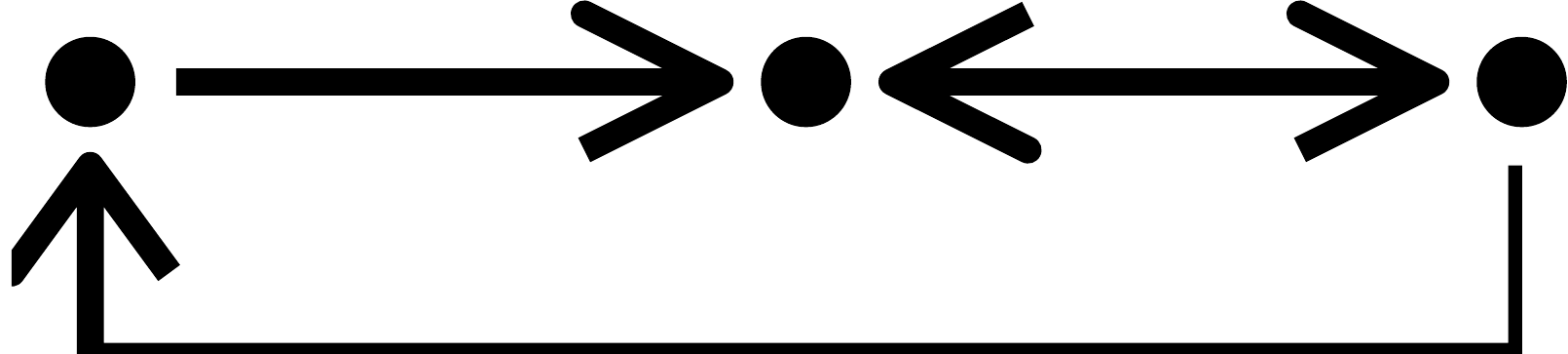}&   &   &   &   &   &   &   &   &   &   &   &   &\mm &   & \mp &  \\
\vspace{-1pt}\hspace{-13pt}\includegraphics[clip,trim=90pt 355pt 59pt 335pt, width=0.08\textwidth]{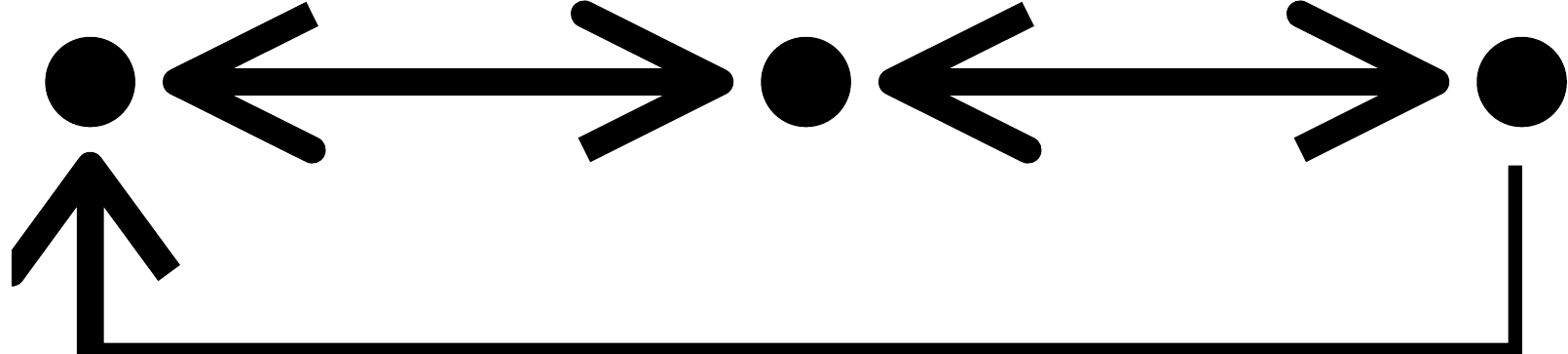}&   &   &   &   &   &   &   &   &   &   &   &   &   &   &\mm & \mp\\
\end{tabular}
\label{AB3}
\end{table}

\subsection{Adaptation of weights: Three-node motifs}

\label{sec_matrix_F}
The adaptation matrix $F\in\R^{N_{\mathrm{mot}}\times N_{\mathrm{mot}}}$ is defined such that $F_{ij}=1$ if motif $j$ can be derived from motif $i$ be adding one edge, and 0 otherwise.
For the three-node motifs, the adaptation matrix is as follows (see Figure \ref{fig1}):
{\begin{center}
$F = $
\begin{tiny}
\begin{tabular}{|cccccccccccccccc|}
0 \ha 1 \ha 0 \ha 0 \ha 0 \ha 0 \ha 0 \ha 0 \ha 0 \ha 0 \ha 0 \ha 0 \ha 0 \ha 0 \ha 0 \ha 0 \\
0 \ha 0 \ha 1 \ha 1 \ha 1 \ha 1 \ha 0 \ha 0 \ha 0 \ha 0 \ha 0 \ha 0 \ha 0 \ha 0 \ha 0 \ha 0 \\
0 \ha 0 \ha 0 \ha 0 \ha 0 \ha 0 \ha 1 \ha 1 \ha 1 \ha 0 \ha 0 \ha 0 \ha 0 \ha 0 \ha 0 \ha 0 \\
0 \ha 0 \ha 0 \ha 0 \ha 0 \ha 0 \ha 1 \ha 0 \ha 1 \ha 0 \ha 0 \ha 0 \ha 0 \ha 0 \ha 0 \ha 0 \\
0 \ha 0 \ha 0 \ha 0 \ha 0 \ha 0 \ha 1 \ha 1 \ha 1 \ha 1 \ha 0 \ha 0 \ha 0 \ha 0 \ha 0 \ha 0 \\
0 \ha 0 \ha 0 \ha 0 \ha 0 \ha 0 \ha 0 \ha 1 \ha 1 \ha 0 \ha 0 \ha 0 \ha 0 \ha 0 \ha 0 \ha 0 \\
0 \ha 0 \ha 0 \ha 0 \ha 0 \ha 0 \ha 0 \ha 0 \ha 0 \ha 0 \ha 1 \ha 1 \ha 1 \ha 0 \ha 0 \ha 0 \\
0 \ha 0 \ha 0 \ha 0 \ha 0 \ha 0 \ha 0 \ha 0 \ha 0 \ha 0 \ha 1 \ha 0 \ha 1 \ha 1 \ha 0 \ha 0 \\
0 \ha 0 \ha 0 \ha 0 \ha 0 \ha 0 \ha 0 \ha 0 \ha 0 \ha 0 \ha 0 \ha 1 \ha 1 \ha 1 \ha 0 \ha 0 \\
0 \ha 0 \ha 0 \ha 0 \ha 0 \ha 0 \ha 0 \ha 0 \ha 0 \ha 0 \ha 0 \ha 0 \ha 1 \ha 0 \ha 0 \ha 0 \\
0 \ha 0 \ha 0 \ha 0 \ha 0 \ha 0 \ha 0 \ha 0 \ha 0 \ha 0 \ha 0 \ha 0 \ha 0 \ha 0 \ha 1 \ha 0 \\
0 \ha 0 \ha 0 \ha 0 \ha 0 \ha 0 \ha 0 \ha 0 \ha 0 \ha 0 \ha 0 \ha 0 \ha 0 \ha 0 \ha 1 \ha 0 \\
0 \ha 0 \ha 0 \ha 0 \ha 0 \ha 0 \ha 0 \ha 0 \ha 0 \ha 0 \ha 0 \ha 0 \ha 0 \ha 0 \ha 1 \ha 0 \\
0 \ha 0 \ha 0 \ha 0 \ha 0 \ha 0 \ha 0 \ha 0 \ha 0 \ha 0 \ha 0 \ha 0 \ha 0 \ha 0 \ha 1 \ha 0 \\
0 \ha 0 \ha 0 \ha 0 \ha 0 \ha 0 \ha 0 \ha 0 \ha 0 \ha 0 \ha 0 \ha 0 \ha 0 \ha 0 \ha 0 \ha 1 \\
0 \ha 0 \ha 0 \ha 0 \ha 0 \ha 0 \ha 0 \ha 0 \ha 0 \ha 0 \ha 0 \ha 0 \ha 0 \ha 0 \ha 0 \ha 0
\end{tabular}
\end{tiny}\end{center}
\vspace{5pt}
Note that $F$ is an upper triangular matrix with zeros in its diagonal, which guarantees the convergence of the series in Eq. (\ref{eq_adapt}).
Therefore, Eq. (\ref{eq_adapt}) can be written as
\begin{equation}w = \left(I-\frac FN\right)^{-1}\tilde w.\label{eq_adapt2}\end{equation}

\subsection{Theoretical network models aiming to promote empty three-node motifs}
\label{sec_theoretical}
In this section, two strategies for connecting $N$ nodes, each with a fixed number ($K\leq\frac N2$) of inputs, in a way that maximizes the number of empty three-node motifs is presented.
For the sake of simplicity of notation, let us assume that $K\geq3$.
Let us call these connection strategies ``intra-connectivity strategy'' and ``inter-connectivity strategy''. In the intra-connectivity strategy, nodes $\{1,...,K+1\}$ are chosen as a subset where each
node connects bidirectionally to each other, and for the rest of the nodes, $\{K+2,...,N\}$, each receives an input from the first $K$ nodes of the previous set, namely, from $\{1,...,K\}$.
Thus, nodes $\{1,...,K\}$ have an output to all nodes in the network, node $K+1$ has an output to the nodes $\{1,...,K\}$, and the nodes $\{K+2,...,N\}$ do not have any outputs.
By contrast, the inter-connectivity strategy is defined so that nodes $\{1,...,K\}$ project to all other nodes, i.e. nodes $\{K+1,...,N\}$, and nodes $\{K+1,...,2K\}$ project back to nodes $\{1,...,K\}$,
  while nodes $\{2K+1,...,N\}$ have no outputs.
Let us show that these two strategies for promoting the empty motifs are sub-optimal in the sense that small changes, namely, changing a source node of a single edge,
does not increase the number of empty motifs. Switching the target node instead of the source node is not possible due to the fixed number of inputs ($K$) each node must have.
\par\bigskip
\textbf{Intra-connectivity strategy}
\par\medskip
The number of empty motifs in the networks following intra-connectivity strategy is determined as follows. The motif formed by three nodes $i_1,i_2,$ and $i_3$ is empty if and only if all three nodes are
outside the intra-connected cluster, namely, $i_1,i_2,i_3 > K+1$, \textbf{or} two of them are outside the intra-connected cluster and the third one is node $K+1$ which does not project
outside the intra-connected cluster. This gives us the number of empty motifs 
$$N_{\mathrm{mot-1}} = \left(\begin{array}{c}N-K-1\\3\end{array}\right) + \left(\begin{array}{c}N-K-1\\2\end{array}\right) = \left(\begin{array}{c}N-K\\3\end{array}\right).$$

Let us show that a change of the source node of a single edge, $j\rightarrow k$ to $j'\rightarrow k$, does not increase the number of empty motifs.
\begin{itemize}
\item{}If $k\leq K$, the source node $j$ is in the intra-connected cluster, $j\in\{1,...,K+1\}\setminus\{k\}$. To change the source node (to a node that does not yet project to $k$),
  we have to pick $j'\in\{K+2,...,N\}$. This change does not alter the number of empty motifs, as one unidirected edge ($k\leftarrow j'$) will be changed to bidirected and one bidirected 
  ($j\leftrightarrow k$) to unidirected.
\item{}If $k=K+1$, the source node $j$ is in the intra-connected cluster, $j\in\{1,...,K+1\}\setminus\{k\}$. To change the source node, similarly as in the previous case, 
  we have to pick $j'\in\{K+2,...,N\}$.
These two nodes ($k$ and $j'$) were part of $N-K-2$ empty motifs that are transformed to motif 2 by the change, but on the other hand, no new empty motifs will be created as there will still be
an edge $k\rightarrow j$ between $j$ and $k$. Hence, the number of empty motifs in the altered graph is
$$N_{\mathrm{mot-1}}' = N_{\mathrm{mot-1}} - (N-K-2).$$
\item{}If $k>K+1$, the source node $j$ is one of the first $K$ nodes, and we pick $j'\in\{K+1,...,N\}\setminus\{k\}$.
  By this operation, $N-K-2$ empty motifs will be transformed to motif 2 (motifs formed by nodes $j'$ and $k$), but on the other hand, $N-K-2$ motifs of type 2 (motifs formed by $j$ and $k$)
    will be transformed to empty motifs, and hence $N_{\mathrm{mot-1}}' = N_{\mathrm{mot-1}}$.
\end{itemize}
Therefore, all possible changes of single edges either reduce or maintain the number of empty motifs. $\square$
\par\bigskip
\textbf{Inter-connectivity strategy}
\par\medskip
The number of empty motifs in the networks following inter-connectivity strategy is determined as follows. The motif formed by three nodes $i_1,i_2,$ and $i_3$ is empty if and only if all three nodes
belong to set $\{1,...,K\}$ \textbf{or} all three belong to set $\{K+1,...,N\}$. This gives us the number of empty motifs 
$$N_{\mathrm{mot-1}} = \left(\begin{array}{c}N-K\\3\end{array}\right) + \left(\begin{array}{c}K\\3\end{array}\right).$$

Let us show that a change of the source node of a single edge, $j\rightarrow k$ to $j'\rightarrow k$, does not increase the number of empty motifs.
\begin{itemize}
\item{}If $k\leq K$, the source node $j$ belongs to the second non-intraconnected set, $j\in\{K+1,...,2K\}$. To change the source node, we have to pick $j'\in\{1,...,K\}\setminus\{k\}\cup\{2K+1,...,N\}$.
  \begin{itemize}
  \item{}If $j'\in\{1,...,K\}\setminus\{k\}$, no empty motifs are created as the link $k\rightarrow j$ remains, but an edge is added among the first non-intraconnected set $\{1,...,K\}$,
  which reduces the number of empty motifs by $K-2$ and leaves us with
$N_{\mathrm{mot-1}}' = N_{\mathrm{mot-1}} - K+2$. 
  \item{}If $j'\in\{2K+1,...,N\}$, the number of empty motifs is not changed as one unidirected edge will be changed to bidirected and one bidirected to unidirected.
  \end{itemize}
\item{}If $K+1\leq k\leq2K$, the source node $j$ is in the set $\{1,...,K\}$. To change the source node, we have to pick $j'\in\{K+1,...,N\}\setminus\{k\}$.
This does not create empty motifs as the link $k\rightarrow j$ remains, but it creates an edge among the non-intraconnected set $\{K+1,...,N\}$,
  reducing the number of empty motifs by $N-K-2$, which leaves us with
$N_{\mathrm{mot-1}}' = N_{\mathrm{mot-1}} - N+K+2$.
\item{}If $k>2K$, the source node $j$ belongs to the first non-intraconnected set, $j\in\{1,...,K\}$, and we have to pick $j'\in\{K+1,...,N\}\setminus\{k\}$.
  By this operation, $N-K-2$ empty motifs will be transformed to motif 2 (motifs formed by nodes $j'$ and $k$), but on the other hand, $(N-2K-1)+(K-1)$ motifs of type 2
  (motifs formed by $j$ and $k$, third node being either in set $\{2K+1,...,N\}\setminus\{k\}$ or in $\{1,...,K\}\setminus\{j\}$) will be transformed to empty motifs, and hence 
  $N_{\mathrm{mot-1}}' = N_{\mathrm{mot-1}} - N+K+2$.  
\end{itemize}
Therefore, all possible changes of single edges either reduce or maintain the number of empty motifs. $\square$

Figure \ref{figSempty} shows the numbers of empty motifs as a function of numbers of inputs $K$ in networks of size $N=1000$ following these two strategies.
Figure \ref{figSempty} also shows that the MBNs promoting empty motifs produce approximately the same number as (although slightly more than) the intra-connectivity strategy.
Until large $K$, the two strategies perform approximately equally well, but for very large $K$ the inter-connectivity strategy produces much more empty motifs than the intra-connectivity
strategy. Notice, however, that the intra-connectivity strategy is possible to use with numbers of inputs larger than $\frac N2$ while inter-connectivity strategy is not.


\subsection{Extension to four-node motifs}

\label{sec_extension_to_four}
While determining the points for node $i$ as a possible input for node $k$ in the case of three-node motifs, it suffices to count the numbers of $N_{\mathrm{premot}}=4^2\times 2$ different
kind of connectivity patterns: 4 possibilities for connection $i$ to $j$, 4 possibilities for connection $j$ to $k$, and 2 possibilities for whether or not there is a connection $k$ to $i$.
By contrast, in the case of four-node motifs, one has to calculate the numbers of $N_{\mathrm{premot}}=4^5\times 2$ different connectivity patterns: 4 possibilities for each of the connections
$i$ to $j_1$, $i$ to $j_2$, $j_1$ to $j_2$, $j_1$ to $k$, and $j_2$ to $k$, and 2 possibilities for whether or not there is a connection $k$ to $i$.
Moreover, due to having to account for two additional nodes ($j_1$ and $j_2$) instead of just one ($j$), there is also a network size dependent difference between the two cases:
As the maximal computational cost of the three-node motif algorithm is in the order of $O(EN^2)$ logical operations ($E$ denoting
the total number of edges), the cost of the four-node motif algorithm is in the order of $O(EN^3)$ operations.

The pseudo-code below shows the function Calculatepoints for four-node motifs. The constant matrices $G$ and $F$ for four-node motif calculations are relatively large (1024$\times$218 and 218$\times$218),
  but they can be determined algorithmically.
\begin{algorithmic}
  \STATE $\lambda=$ Calculatepoints\_4($k$,$M$):
  \FOR {node index $i\neq k$ s.t. $M_{ik}=0$}
   \FOR {$r\in\{1,\ldots,{N_{\mathrm{premot}}}\}$}
    \STATE Set the number of existing pre-motifs to zero:\\ \hspace{10pt}$Q_{ir}\leftarrow 0$
    \FOR {$j_1\in\{1,\ldots,N\}\backslash\{i,k\}$}
     \FOR {$j_2\in\{1,\ldots,N\}\backslash\{i,k,j_1\}$}
      \IF {the connections between nodes $i$, $j_1$, $j_2$, $k$ form the pre-motif $r$}
       \STATE Set $Q_{ir}\leftarrow Q_{ir}+1$
      \ENDIF
     \ENDFOR
    \ENDFOR
   \ENDFOR
   \STATE Count the points for node $i$ as \\ \hspace{10pt}$\lambda_i \leftarrow\sum_{r=1}^{N_{\mathrm{premot}}}\sum_{m=1}^{N_{\mathrm{mot}}}Q_{ir}G_{rm}w_m$
  \ENDFOR
  \STATE Return vector $\lambda$
\end{algorithmic}

\subsection{Calculation of clustering coefficient}
\label{sec_clustering_coefficient}
The clustering coefficient of a graph $M$ is calculated as the average over local clustering coefficients $C_i(M)$, $i\in\{1,\ldots,N\}$,
where $C_i(M)$ is defined as the number of connected triangles in the neighbourhood of node $i$ \cite{watts1998collective}. When counting the triangles,
the directions of the edges are relaxed in the way that they need not be traversable, yet changing a unidirected edge of a connected triple to bidirected doubles the number
of counted triangles. In this work, the formula \cite{maki-marttunen2013structure}
\begin{equation}C_i(M)\hspace{-2pt}=\hspace{-2pt}\frac 1{8\left(\begin{array}{c}m_i\\2\end{array}\right)}\hspace{-6pt}\displaystyle\sum_{\begin{array}{c}j=1\\j\neq i\end{array}}^N
\hspace{-10pt}\sum_{\begin{array}{c}k=1\\k\neq i\end{array}}^{j-1}\hspace{-2pt}(M_{ij}+M_{ji})\cdot(M_{ik}+M_{ki})\cdot(M_{jk}+M_{kj})
\label{localclusteringcoefficient}
\end{equation}
is used for determining the local clustering coefficient of node $i$, where
$m_i$ is the number of neighbours of node $i$, i.e. the number of nodes that node $i$ projects to or receives input from.


\subsection{Calculation of modularity}
\label{sec_modularity}
The modularity of a graph $M$ is calculated in this work using Eq. (\ref{modularity}).
For obtaining the communities, the hierarchical clustering is applied, where the Hamming distance in the output and input node patterns is used as the determining factor for joining two clusters.
In more detail, the distance between two nodes, $i$ and $j$, is defined as
\begin{equation}d_{ij}=\displaystyle\frac 12\sum_{\begin{array}{c}k=1\\i\neq k\neq j\end{array}}^N(|M_{ik}-M_{jk}|+|M_{ki}-M_{kj}|),\end{equation}
and when joining two nodes $i_1$ and $i_2$, the distance of the new node
$i_3$ to the other nodes $j$ in the network is determined iteratively as
\begin{equation}d_{i_3j} = \frac 12d_{i_1j}+\frac 12d_{i_2j}.\end{equation}
For reference, an alternative partition method \cite{zhou2005learning} is used in the calculation of modularity in Figure \ref{fig4}E. This method starts with a
giant population-wide cluster, and divides it iteratively into two.
With each iteration, the division of each cluster into two, as described in \cite{zhou2005learning}, is tested. Then, the modularity of each proposal clustering is
calculated, and the division that produced the largest modularity value is applied, while the other divisions are ignored. This is repeated until the desired number of clusters is obtained.

The main functional difference between
these two clustering methods is that the latter assumes the modules to have a large intra-cluster connectivity, while the former also accepts more unconventional types of modularity. As an example, a
network where nodes in cluster A project densely to nodes in cluster B, and vice versa, while both are sparse in intra-cluster connections, would be easily divided to modules A and B by the applied
hierarchical clustering method, but this is not the case with the clustering method of \cite{zhou2005learning}. Another difference between the methods is that the former starts by assigning each node
in its own cluster and progresses by combining these clusters, while the latter starts with just one giant cluster and cuts it into two clusters until a given number of clusters is attained --- although
the authors of \cite{zhou2005learning} state that their framework also allows a direct clustering into any number of clusters instead of the iterative splitting into two.


While the same measure of modularity is largely used in the research of undirected networks, there are several options for the measure of modularity for directed networks. Figure \ref{fig4}
shows the results obtained using a modification of \cite{newman2006modularity} for directed networks (Eq. \ref{modularity}), implemented as in \cite{arenas2007size}.
Nevertheless, the results stay qualitatively similar
if a slightly simpler formula of modularity, where it is approximated that $\frac{m_i^{\mathrm{out}}m_j^{\mathrm{in}}}E\approx p$, is applied, as shown in Figure \ref{figS5}.

\subsection{Parameter optimization}
\label{sec_parameter_optimization}
The MBN algorithm weights can be successfully optimized to promote certain global network measures such as small-worldness and modularity, as shown in Figure \ref{fig4}. For this, a genetic optimization
algorithm was used. During each function evaluation, 20 networks of smaller size were generated for each desired in-degree distribution. The function output was chosen as the average small-worldness index or
the average modularity of the formed networks, and this function was maximized over possible values of weight $\tilde w$.

In the maximization of small-worldness index, the in-degree distributions were chosen as delta-weighted distributions $p_{\mathrm{in}}(d)=\delta_K(d)$, $K=2,\ldots,6$. Weights were chosen as
$\tilde w=(\alpha_1,0,0,\alpha_2,0,0,0,0,0,0,0, \alpha_3,0,0,0,\alpha_4)$, and the optimal weights were obtained with $\alpha=(-1.351,1.407,1.755,0.567)$. Optimization of these four weights only,
which correspond to the motifs that only contain bidirected edges, was found adequate to promote the neighbourhood structure needed for the high values of small-worldness.
In the maximization of the modularity, the in-degree distributions were binomial with $p=1/N_{\mathrm{clust}}$, $N_{\mathrm{clust}}=2,\ldots,20$, and the modularity was calculated using hierarchical
clustering into $N_{\mathrm{clust}}$ communities.
Weights were chosen as $\tilde w=(\alpha_1,\alpha_2,0,\alpha_3,0,0,0,0,\alpha_4,0,0,\alpha_5,0,0,0,\alpha_6)$, and the optimal weights were
obtained with $\alpha=(1.852, 1.300,$ $0.838, 0.084, -2.111, 0.1317)$.
Here, the weights of the motifs 2 (a single connection) and 9 (input from a fully connected pair) were varied in addition to those used in the optimization of the small-worldness
index in order to allow the formation of the community structure needed for high values of modularity.
A nearly as good promotion of modularity was found when varying weights of motifs 6 (divergent connection) and 7 (input to fully connected pair) instead of motifs 2 and 9 (data not shown).
The network size was chosen as $N=100$ in the maximization of small-worldness, and as $N=60$ in the maximization of modularity. The MATLAB
genetic algorithm \texttt{ga} (unconstrained) was used with default parameters for solving the optimization problem.

As the obtained weights showed the desired behaviour for networks larger than the ones used during the optimization ($N=200$, see Figure \ref{fig4}B, D and E), it is expected that they could be applicable to
other network sizes as well. This could be a great facilitation to problems, where one wants to implement certain network characteristics in large networks, but due to computational costs is only able to do the
optimization using smaller networks. However, it is not yet certain which global graph properties are conserved between different network sizes in the MBN framework. Exploring this is left for future work.

\begin{figure}[hb!]
\includegraphics[clip,trim=68pt 60pt 155pt 555pt, width=0.95\textwidth]{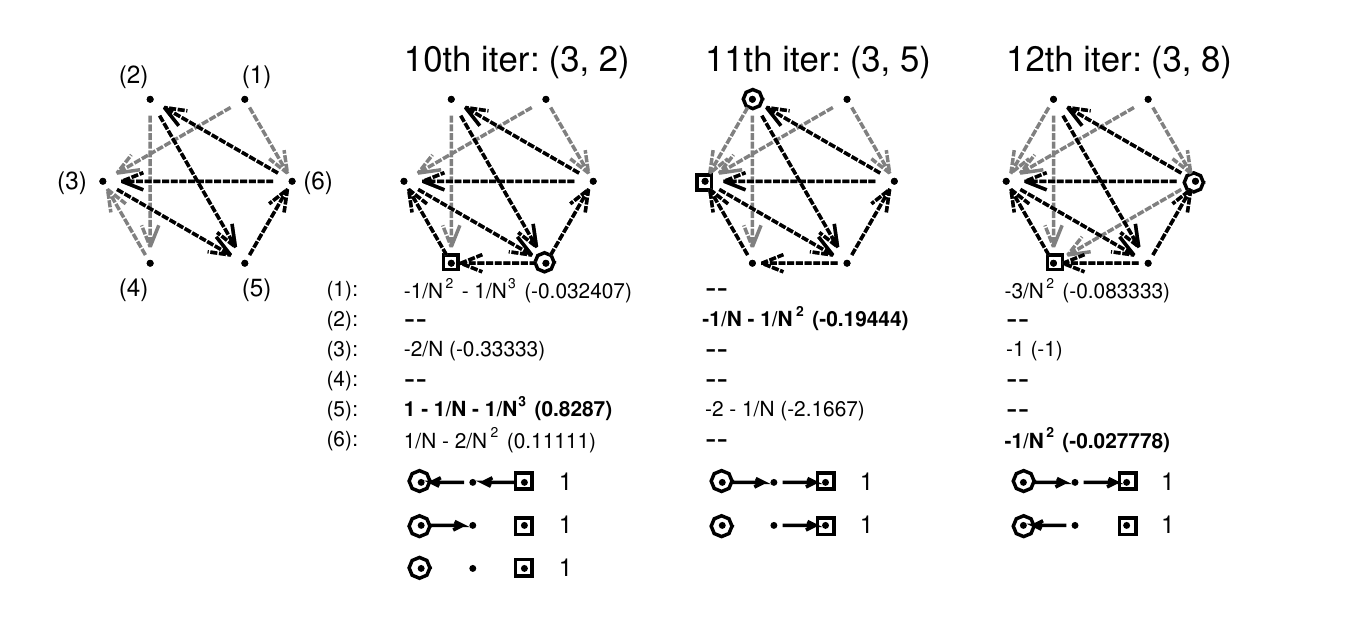}
\caption{Three iterations of the MBN algorithm pronouncing motif 10 (FB motif), network size $N=6$. The weights are chosen as $\tilde{w}_i=\delta_{10}(i)$,
and they are adapted using Eq. (\ref{eq_adapt}). The numbers of FB and FF motifs (\#motifs 10, \#motifs 8) are printed beside the
iteration number. 
The edges that are part of at least one FB motif are black, while others are gray.
The in-degree distribution was binomial with $p=0.33$. 
The 11th and 12th iterations show that due to the density of edges, the number of FB motifs cannot be increased by any choices of input node.
Instead, the choices of inputs that retain the already formed FB motifs induce a great increase in the numbers of FF motifs, as shown by the
rapid increase from 2 to 8 in \#motifs 8.
}
\label{figS0}
\end{figure}

\begin{figure}[hb!]
\centering
\includegraphics[clip,trim=27pt 10pt 25pt 350pt, width=0.95\textwidth]{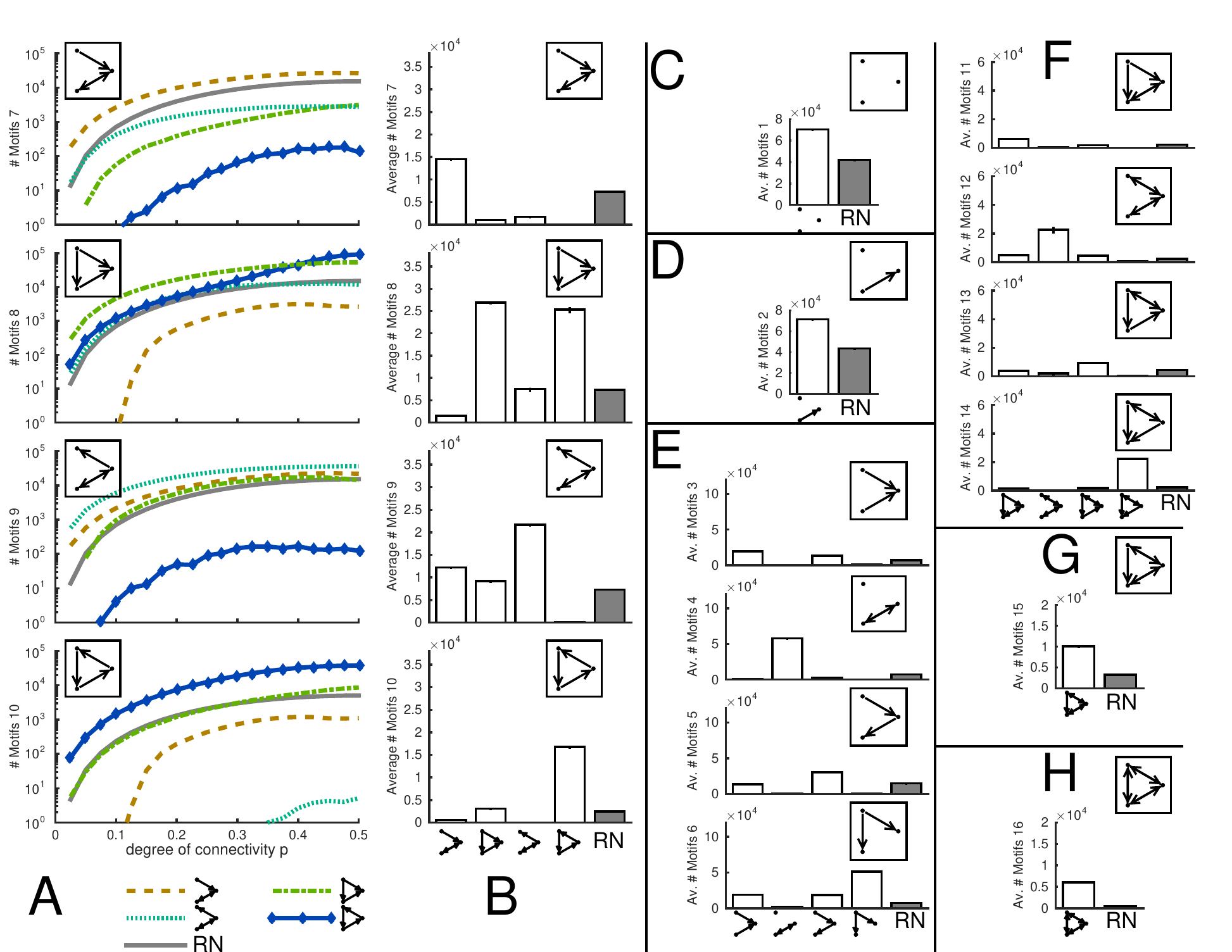}
\caption{\small The networks produced with the MBN algorithm, pronouncing a single three-node motif ($\tilde w=\delta_i$, $i=1,...,16$), show a significant
increase in the numbers of the corresponding motif compared to other networks. \textbf{A}: The numbers of different three-node motifs with three edges as a function of connection probability
$p$. Network size is $N=100$, and binomial in-degree distribution is used. The different curves show the numbers of the considered motif when different weights are used:
$\tilde w=\delta_7$,$\delta_8$,$\delta_9$,$\delta_{10}$, or $0\in\R^{16}$ (random network, denoted by RN). The four panels show the numbers of four different motifs.
The values of the curves represent sample averages of $\Nsamp=200$ realizations.
\textbf{B}: The numbers of the three-node-three-edge motifs shown in (A), averaged (integrated and multiplied by two) over connection
probabilities $0<p\leq 0.5$. The bar height shows the mean of $\Nsamp=200$ realizations and the tick on the bar shows the standard deviation (when invisible, the standard deviation was very small).
\textbf{C--H}: The analysis of (B) repeated for three-node motifs with 0 (C), 1 (D), 2 (E), 4 (F), 5 (G), or 6 (H) edges.
Comparisons are made among MBNs pronouncing another motifs with the same number of edges, and with random networks. The medians of the motif count distributions
are always significantly higher in those MBNs that pronounce the corresponding motif than in any other shown network (U-test, $p=0.05$).
}
\label{figS1}
\end{figure}

\begin{figure}[hb!]
\centering
\includegraphics[clip,trim=17pt 20pt 55pt 0pt, width=0.95\textwidth]{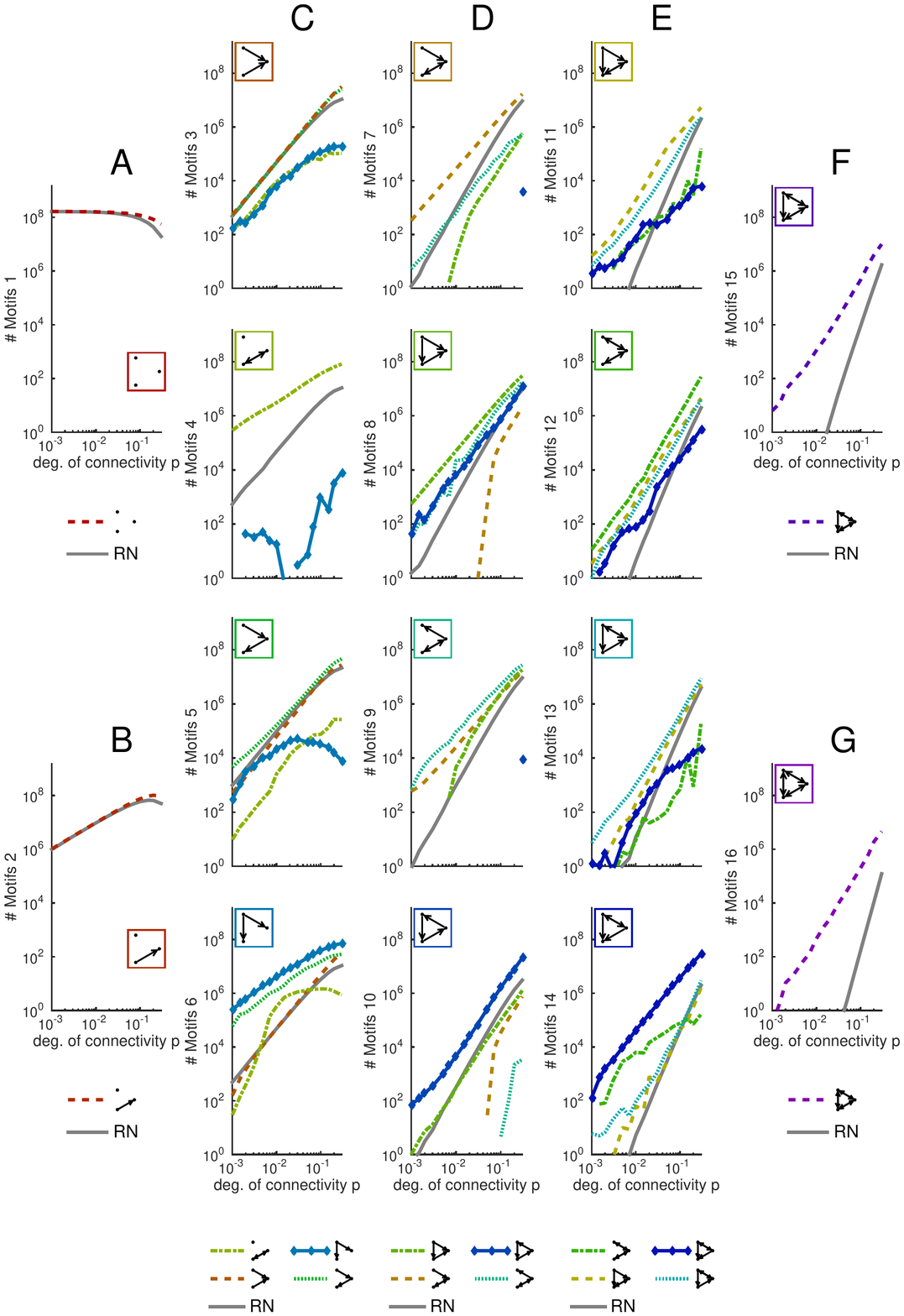}
\caption{\small The large, single motif-promoting MBNs show an increase in the numbers of the corresponding motif compared to other networks.
The curves show the numbers of motifs in large ($N$=1000) networks. Each MBN promoting the considered motif is compared to random networks and other MBNs promoting another motifs with the same number
of edges, when possible. All compared networks have the same in-degree distribution (binomial, $p\in[0.001,0.3]$).
In panels (A)--(G), the MBNs promoting empty motifs (A), motifs of one (B), two (C), three (D), four (E), or five (F) edges, or fully connected motifs (G) are shown.
The values of the curves represent sample averages of $\Nsamp=3$ realizations.
The MBN promoting the considered motif always gave more motifs of that kind than other MBNs (MBNs promoting another motif with the same number of edges) or RNs, but 
for some motifs, the differences between the best and second best MBN were not large, see e.g. \#motifs 3 in the MBN promoting motif 3 (\#motifs 3 in range [549,$3.17\times10^7$] for
$p\in[0.001,0.3]$) versus the MBN promoting motif 5 (\#motifs 3 in range [531,$2.57\times10^7$]) in panel C.
}
\label{figS2}
\end{figure}

\begin{figure}[hb!]
\centering
\includegraphics[clip,trim=0pt 20pt 90pt 360pt, width=0.95\textwidth]{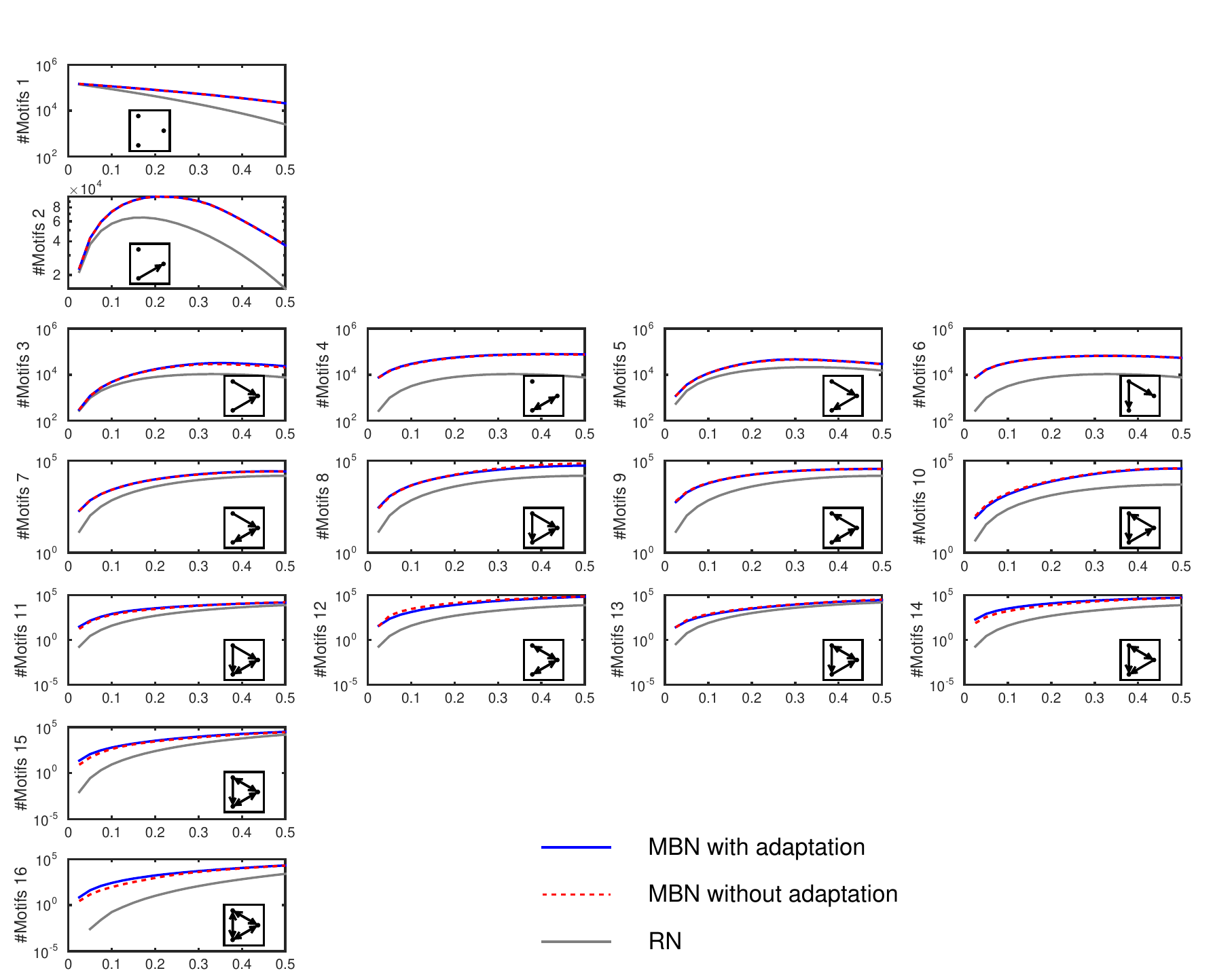}
\caption{\small 
Comparison of motif counts in MBNs promoting the considered motif with (solid blue) or without (dashed red) weight adaptation.
Comparison made to random networks (gray) as well. Network size $N$=100, curves represent sample means of $\Nsamp=200$ realizations.
For densely connected motifs (motifs 15 and 16), the MBN using weight adaptation produces higher numbers of motifs than MBN without adaptation.
For sparsely connected motifs (motifs 1 and 2) there is no visible difference, and for motifs of 2, 3 or 4 edges, the outcome varies between the two strategies.
}
\label{figS3}
\end{figure}

\begin{figure}[hb!]
\includegraphics[clip,trim=25pt 180pt 80pt 365pt, width=0.95\textwidth]{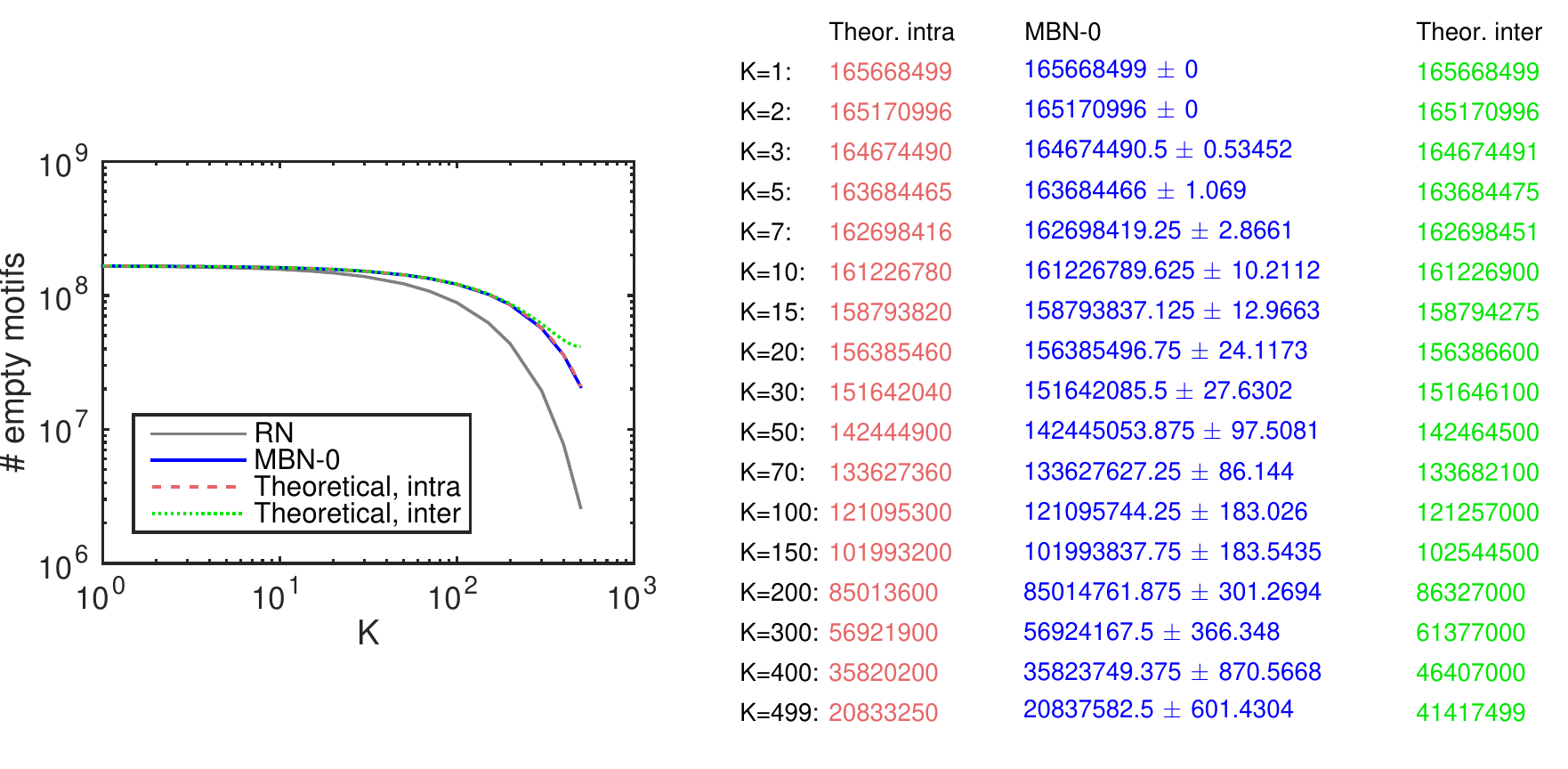}
\caption{\small 
The MBNs promoting empty motifs contain a number of empty motifs that lies between those contained by networks following the intra- and inter-connectivity strategies
  (see Supplementary material Section \ref{sec_theoretical}).
The y axis shows the number of empty motifs in networks of size $N=1000$, and x axis shows the number of inputs $K$.
Solid blue line shows results for the MBN promoting empty motifs, while red and green dashed line show results for the intra- and inter-connectivity strategies, respectively,
and solid gray line shows the results for random networks. On the right-hand side of the figure, the exact numbers are given (for MBNs, the sample mean and standard
deviation are given, $N_{\mathrm{samp}}=3$).
}
\label{figSempty}
\end{figure}

\begin{figure}[hb!]
\includegraphics[clip,trim=0pt 0pt 45pt 360pt, width=0.95\textwidth]{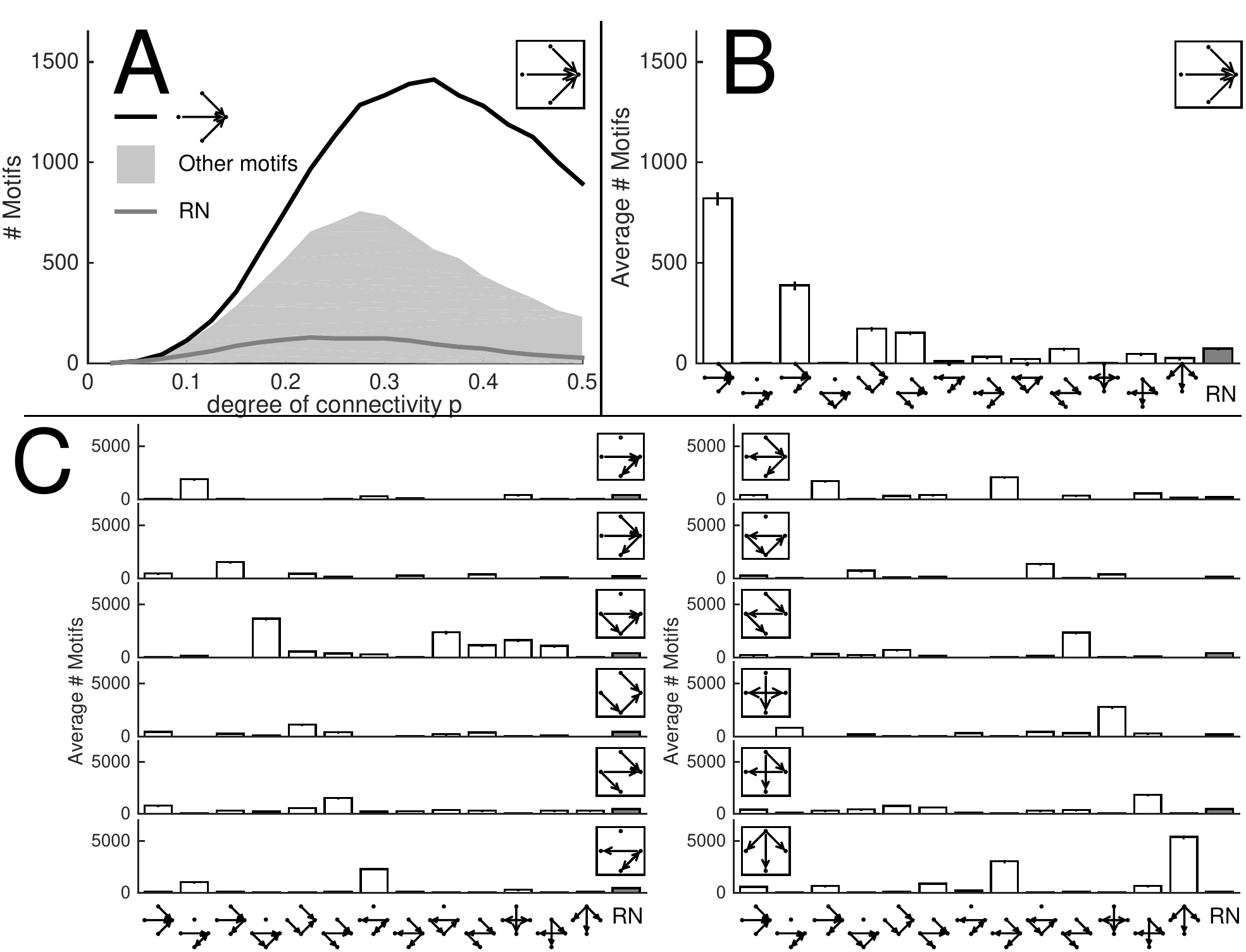}
\caption{\small The MBNs that pronounce a single four-node motif ($\tilde w=\delta_i$, $i=1,...,218$)
possess significantly larger fraction of the stressed motif than other networks do.    
\textbf{A}: The number of a certain four-node-three-edge motif as a function of connection probability (see Figure \ref{fig3}A). The average number of motifs in the MBN promoting the considered
motif is shown with the black curve, while the corresponding curves of MBNs promoting other four-node-three-edge motifs fall in the gray area.
The network size is $N=30$, binomial in-degree distribution is used, and the shown data are averages of $\Nsamp=100$ samples.
\textbf{B}: The numbers of the four-node-three-edge motif in the MBNs shown in panel A, averaged over connection probabilities $0<p\leq 0.5$. See Figure \ref{fig3}C-D for
details. \textbf{C}: The analysis of panel B repeated for other four-node-three-edge motifs. The medians of the motif count distributions
are always significantly higher in those MBNs that pronounce the corresponding motif than in any other shown network (U-test, $p=0.05$).
}
\label{figS4}
\end{figure}

\begin{figure}
\includegraphics[clip,trim=270pt 170pt 10pt 290pt, width=0.75\textwidth]{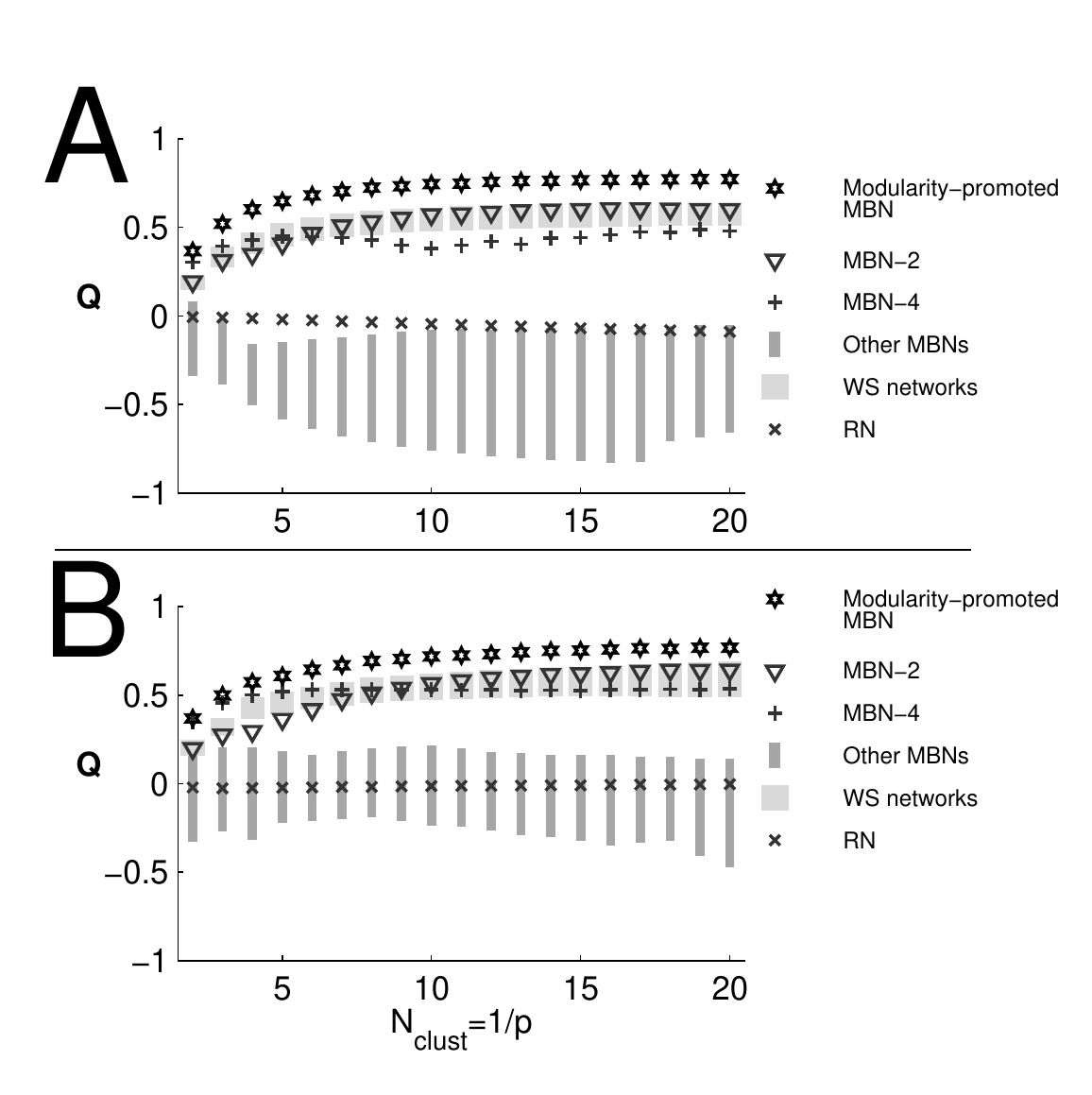}
\caption{
Modularities of different networks when a simpler formula of modularity is used. Here, the modularity $Q$ is
calculated as
$Q(M) = \frac 1E\sum_{\mathrm{cluster}\ {\cal I}}\sum_{i,j\in {\cal I}}(M_{ij} - p)$, where $p$ is the mean
connection probability of the network. The panel \textbf{A} shows the results when hierarchical clustering is
used (see Figure \ref{fig4}D for corresponding results obtained with Eq. (\ref{modularity})) and the panel \textbf{B} shows the results obtained with iterated Zhou's clustering
method (see Figure \ref{fig4}E).
}
\label{figS5}
\end{figure}

\begin{figure}
\includegraphics[clip,trim=0pt 0pt 30pt 340pt, width=0.95\textwidth]{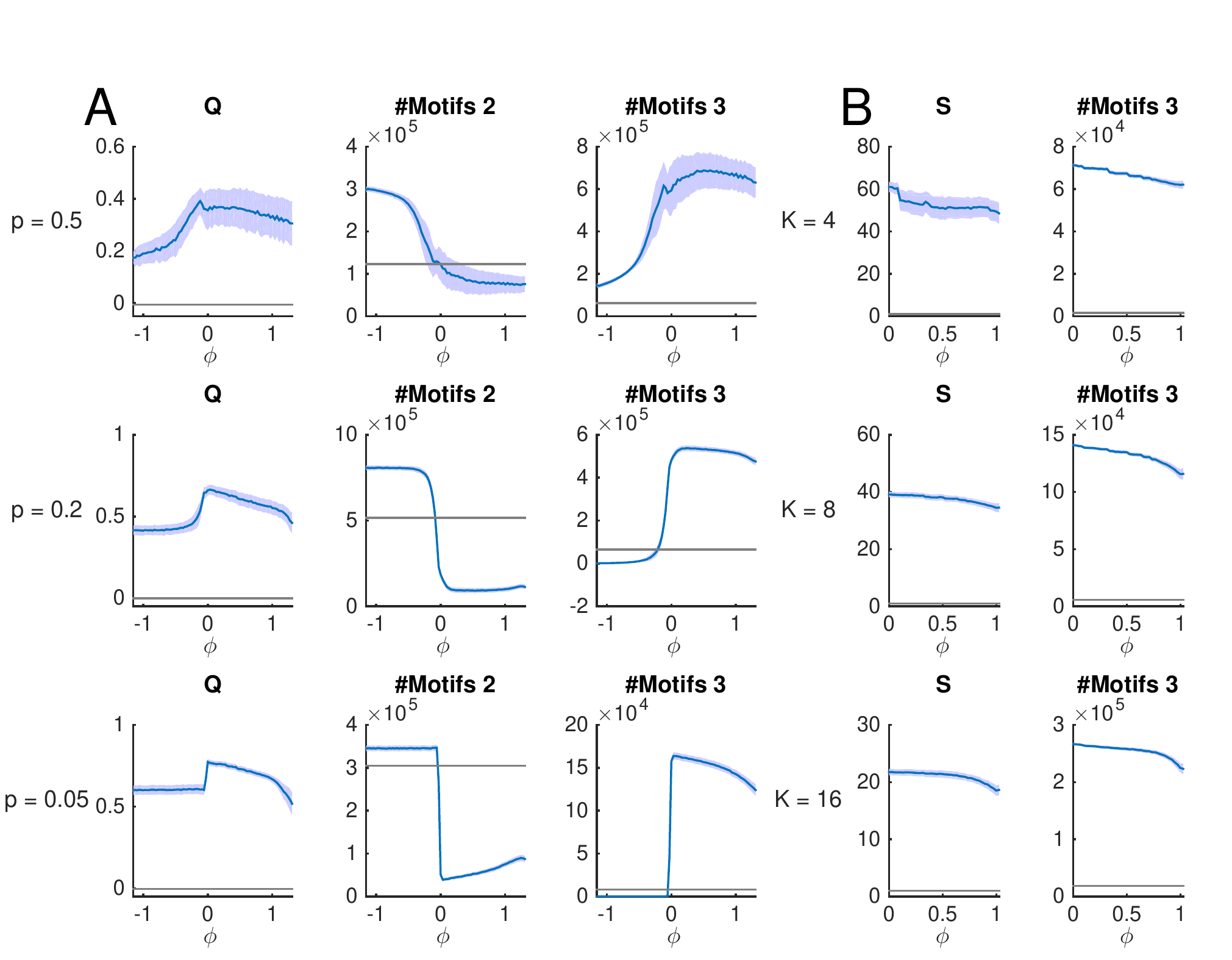}
\caption{
A continuum of networks from single motif-promoting MBNs to small-worldness- or modularity-optimized MBNs.
\textbf{A}: Continuum from modularity-optimized MBNs (weight vector
$\tilde w=(1.852,1.3,0,0.838,0,0,0,0,0.084,0,0,-2.111,0,0,0,0.1317)$, as in Figure \ref{fig4})
to single motif-promoting MBNs.
The continuum is parametrized by the angle 
$\phi$ between the modularity-optimized MBN and the considered MBN, whose weight vector lies on the unit sphere arc from the single motif-promoting MBN $\tilde w=\delta_2$
(negative values of $\phi$) or $\tilde w=\delta_3$ (positive values of $\phi$) to the unit vector corresponding to the modularity-optimized weight vector.
The angle between two weight vectors, $\tilde w^{(1)}$ and $\tilde w^{(2)}$ is calculated as 
$\phi=\mathrm{arccos} \frac{\tilde w^{(1)}\cdot\tilde w^{(2)}}{||\tilde w^{(1)}|| ||\tilde w^{(2)}||}$.
Hence, the modularity-optimized MBNs reside at $\phi=0$, the single motif-promoting MBNs are at the two ends of the shown range,
and the continuum is represented by 38 intermediate weight vectors on both sides of the modularity-optimized MBNs.
The left-hand panels show the modularity values, and the middle and right-hand panels show the numbers
of motifs 2 (middle) and 3 (right). Blue curves and the shaded areas show the sample mean and standard deviation of the corresponding MBNs ($N_{samp}=500$),
and the gray line shows the average value of the RNs with the same in-degree distribution.
The upper panels show the results for dense connectivity ($p=0.5$, clustering made to two modules in the calculation of $Q$),
while the middle panels for medium connectivity ($p=0.2$, clustering made to five modules), and the lower panels for sparse connectivity ($p=0.05$, clustering made to twenty modules).
\textbf{B}: Continuum from small-worldness-optimized MBNs (weight vector $\tilde w=(-1.351,0,0,1.407,0,0,0,0,0,0,0,1.755,0,0,0,0.567)$,
as in Figure \ref{fig4}, located at $\phi=0$) to MBNs with weight vector $W=\delta_3$.
The left-hand panels show the small-worldness indices, and right-hand panels show the numbers
of motifs 3. 
The upper panels show the results for networks with $K=4$ (each node receives exactly $K=4$ inputs),
while the middle and lowest panels show the results for networks with $K=8$ and $K=16$.
Although all shown MBNs express high values of modularity (A) or small-worldness (B) compared to RNs, there is a regime near $\phi=0$ where these values are higher.
At some connection probabilities, a non-zero $\phi$ gives yet larger values of $Q$ or $S$, but the optimized MBNs ($\phi=0$) perform well across the connectivities.
}
\label{figS6}
\end{figure}

\end{document}